\magnification=\magstephalf
\baselineskip=16pt
\parskip=8pt
\rightskip=0.5truecm

\def\e{\epsilon}

\def\g{\gamma}

\def\s{\sigma}

\def\x{\xi}

\def\D{\Delta}
\def\L{\Lambda}

\def\O{\Omega}

\def\del #1{\frac{\partial^{#1}}{\partial\l^{#1}}}

\def\1{1}

\def\E{I\kern-.25em{E}}
\def\N{I\kern-.22em{N}}
\def\M{I\kern-.22em{M}}
\def\R{I\kern-.22em{R}}
\def\Q{I\kern-.22em{Q}}
\def\Z{{Z\kern-.5em{Z}}}

\def\C{C\kern-.75em{C}}
\def\P{I\kern-.25em{P}}

\def\del{\partial}


\def\EE{{\cal E}}

\def\GG{{\cal G}}
\def\HH{{\cal H}}

\def\chap #1{\line{\ch #1\hfill}}

\def\one{\hbox{J}\kern-.2em\hbox{I}}

\def\ov #1{\overline{#1}}
\def\ba{{\backslash}}
\def\sb{{\subset}}
\def\sp{{\supset}}

\def\em{{\emptyset}}


\newcount\foot
\foot=1
\def\note#1{\footnote{${}^{\number\foot}$}{\ftn #1}\advance\foot by 1}
\def\tag #1{\eqno{\hbox{\rm(#1)}}}
\def\frac#1#2{{#1\over #2}}
\def\text#1{\quad{\hbox{#1}}\quad}

\def\proposition #1{\noindent{\thbf Proposition #1: }}
\def\datei #1{\headline{\rm \the\day.\the\month.\the\year{}\hfill{#1.tex}}}

\def\theo #1{\noindent{\thbf Theorem #1: }}
\def\lemma #1{\noindent{\thbf Lemma #1: }}

\def\corollary {\noindent{\thbf Corollary: }}
\def\proof{{\noindent\pr Proof: }}
\def\proofof #1{{\noindent\pr Proof of #1: }}
\def\endproof{$\diamondsuit$}
\def\remark{{\bf Remark: }}

\def\endproof{$\diamondsuit$}

\font\pr=cmbxsl10 scaled\magstephalf
\font\thbf=cmbxsl10 scaled\magstephalf

\long\def\fussnote#1#2{{\baselineskip=10pt
\setbox\strutbox=\hbox{\vrule height 7pt depth 2pt width 0pt}
\sevenrm
\footnote{#1}{#2}
}}


\font\ch=cmbx12

\font\ftn=cmr8

\font\it=cmti10
\font\bf=cmbx10
\font\srm=cmr5


\overfullrule=0pt
\font\tit=cmbx12
\font\aut=cmbx12
\font\aff=cmsl12
{$  $}
\vskip2truecm
\centerline{\tit 
(NON-) GIBBSIANNESS AND PHASE TRANSITIONS
}
\vskip.2truecm
\centerline{\tit IN RANDOM LATTICE SPIN MODELS
\footnote{${}^*$}{\ftn Work
supported by the DFG
Schwerpunkt `Wechselwirkende stochastische Systeme hoher Komplexit\"at'
}}
\vskip2truecm
\vskip.5truecm
\centerline{\aut  Christof K\"ulske\footnote{${}^{1}$}{\ftn
e-mail: kuelske@wias-berlin.de}
}


\vskip.1truecm
\centerline{\aff WIAS}
\centerline{\aff Mohrenstrasse 39}
\centerline{\aff D-10117 Berlin, Germany}
\vskip1truecm\rm

\noindent {\bf Abstract:}
We consider disordered lattice spin models
with finite volume Gibbs measures $\mu_{\L}[\eta](d\s)$.
Here $\s$ denotes a lattice spin-variable and 
$\eta$ a lattice random variable with 
product distribution $\P$ describing the disorder of the model. 
We ask: When will the joint measures
$\lim_{\L\uparrow\Z^d}\P(d\eta)\mu_{\L}[\eta](d\s)$
be [non-] Gibbsian measures
on the product of spin-space and disorder-space?
We obtain general criteria for both Gibbsianness 
and non-Gibbsianness 
providing an interesting link between phase transitions
at a fixed random configuration and Gibbsianness
in product space: 
Loosely speaking, a phase transition 
can lead to non-Gibbsianness, (only) if
it can be observed on the spin-observable conjugate to the 
independent disorder variables. 

Our main specific example is the random field Ising model
in any dimension for which we show almost sure- 
[almost sure non-] Gibbsianness for the single- [multi-] phase region.  
We also discuss models with disordered couplings, including 
spinglasses and ferromagnets, 
where various mechanisms are responsible for 
[non-] Gibbsianness.


\vfill
     ${}$
\eject



\magnification=\magstephalf
\baselineskip=16pt
\parskip=8pt
\rightskip=0.5truecm

\def\e{\epsilon}

\def\g{\gamma}

\def\s{\sigma}

\def\x{\xi}

\def\D{\Delta}
\def\L{\Lambda}

\def\O{\Omega}

\def\del #1{\frac{\partial^{#1}}{\partial\l^{#1}}}

\def\1{1}

\def\E{I\kern-.25em{E}}
\def\N{I\kern-.22em{N}}
\def\M{I\kern-.22em{M}}
\def\R{I\kern-.22em{R}}
\def\Z{{Z\kern-.5em{Z}}}

\def\C{C\kern-.75em{C}}
\def\P{I\kern-.25em{P}}

\def\del{\partial}


\def\EE{{\cal E}}

\def\GG{{\cal G}}
\def\HH{{\cal H}}

\def\chap #1{\line{\ch #1\hfill}}

\def\one{\hbox{J}\kern-.2em\hbox{I}}

\def\ov #1{\overline{#1}}
\def\ba{{\backslash}}
\def\sb{{\subset}}
\def\sp{{\supset}}

\def\em{{\emptyset}}


\newcount\foot
\foot=1
\def\note#1{\footnote{${}^{\number\foot}$}{\ftn #1}\advance\foot by 1}
\def\tag #1{\eqno{\hbox{\rm(#1)}}}
\def\frac#1#2{{#1\over #2}}
\def\text#1{\quad{\hbox{#1}}\quad}

\def\proposition #1{\noindent{\thbf Proposition #1: }}
\def\datei #1{\headline{\rm \the\day.\the\month.\the\year{}\hfill{#1.tex}}}

\def\theo #1{\noindent{\thbf Theorem #1: }}
\def\lemma #1{\noindent{\thbf Lemma #1: }}

\def\corollary {\noindent{\thbf Corollary: }}
\def\proof{{\noindent\pr Proof: }}
\def\proofof #1{{\noindent\pr Proof of #1: }}
\def\endproof{$\diamondsuit$}
\def\remark{{\bf Remark: }}

\def\endproof{$\diamondsuit$}

\font\pr=cmbxsl10 scaled\magstephalf
\font\thbf=cmbxsl10 scaled\magstephalf

\long\def\fussnote#1#2{{\baselineskip=10pt
\setbox\strutbox=\hbox{\vrule height 7pt depth 2pt width 0pt}
\sevenrm
\footnote{#1}{#2}
}}


\font\ch=cmbx12

\font\ftn=cmr8

\font\it=cmti10
\font\bf=cmbx10
\font\srm=cmr5




\chap{I. Introduction}

The purpose of this paper is to present a
class of measures on discrete lattice spins
showing a rich behavior w.r.t. their Gibbsianness
properties. The examples we consider turn
up in a natural context of well-studied
disordered systems.

Given a random lattice system, such as the random field
Ising model, we look at the 
{\bf joint} distribution of spins and 
random variables describing the disorder. 
It is now very natural from a probabilistic point of 
view to consider the corresponding {\bf joint measures}
on the skew space resulting from the a-priori 
distribution of the disorder variables. 
Taking the infinite volume limit leads
to infinite volume measures on the skew space. 
We will investigate the Gibbsianness-properties
of such measures, for general finite range potentials. 
As we will see, this gives rise to 
a whole family of interesting examples of measures
with non-trivial behavior.

Why consider these measures?- 
Gibbs measures are the basic objects 
for a mathematically rigorous description 
of equilibrium statistical mechanics.
They are characterized by the fact that their
finite volume conditional expectations
can be written in terms of an absolutely summable 
interaction potential. 
The failure of the Gibbsian property 
is linked to the emergence of long-range correlations
or hidden phase transitions. 

In the theory of disordered systems on the other hand, 
the understanding of potentially non-local 
behavior as a function 
of the disorder variables is very important. 
It is a general theme that comes up very soon in any serious analysis 
of a lot of disordered systems. E.g., 
it leads to technically involved concepts
like that of a `bad region' in space where the
realization of the random variable was exceptional 
that must be treated carefully because it could lead to non-locality. 

Now, as we will see in our general investigation, 
the [non-] Gibbsianness of the joint measures is related in an interesting
way to the [non-] locality of certain expectations 
of random Gibbs-measures as a 
function of the disorder variables.  
Since such a non-locality can arise
in a variety of different ways, there is 
a variety of different `mechanisms' for non-Gibbsianness. 
So, the much-disputed phenomenon of non-Gibbsianness
becomes related in a somewhat surprising way  
to continuity questions of the random
Gibbs measures on the spins w.r.t. disorder, 
or, in other words, phase transitions
induced by changes of the disorder variables.  

The present investigation was motivated by the special
recent example of the Ising-ferromagnet with site-dilution
(`GriSing random field')
that was shown to be non-Gibbsian but almost Gibbsian in 
[EMSS] where an interesting realization of the disorder
variables leading to `non-continuity' was found. 
Mathematically the analysis was simplified here 
because the system considered breaks down into finite pieces.
This is of course not true in most of the 
systems of interest (say: the  
random field Ising model). Such a `non-decoupling'
is going to be an essential complication of the 
general treatment we are going to present, 
as we will see. 

Let us remark that there has been some discussion during the last 
years about numerous examples of non-Gibbsian measures,
to what extent the failure of the Gibbsian property
has to be taken serious, and what suitable generalizations
of Gibbsianness should be 
(see e.g. [F],[E],[DS],[BKL],[MRM], references therin,
and the basic paper [EFS]). 
While this discussion still does not seem to be finished,
the answers seem to depend on the specific situation. 
Our point in this context is less a general philosophical one,
but to provide interesting examples that show
(non-)Gibbsianness in a slightly different light 
related to important issues 
in the theory of random Gibbs measures.

More precisely we will do the following: 

\bigskip 

\line{\bf Basic Definitions:\hfill}

Denote by $\O=\O_{0}^{\Z^d}$
the space of {\bf spin-configurations} 
$\s=\left(\s_x \right)_{x\in \Z^d}$, 
where $\O_0$ is a finite set. 
Similarly we denote by $\HH=\HH_{0}^{\Z^d}$ 
the space of {\bf disorder variables}
$\eta=\left(\eta_x \right)_{x\in \Z^d}$
entering the model, where $\HH_0$ is a finite set.
Each copy of $\HH_0$ carries a measure $\nu(d\eta_x)$
and $\HH$ carries the product-measure over the sites,
$\P=\nu^{\otimes_{\Z^d}}$. We denote the 
corresponding expectation by $\E$. 
The space of joint configurations $\O\times\HH=
\left(\O_{0}\times \HH_{0} \right)^{\Z^d}$ 
is called {\bf skew space}. It is equipped with
the product topology. 

We consider disordered 
models whose formal {\bf infinite volume 
Hamiltonian} can be written in terms of  
terms of disordered potentials $(\Phi_A)_{A\sb \Z^d}$,
$$
\eqalign{
&H^{\eta}(\s)=\sum_{A\sb \Z^d}\Phi_{A}\left(\s,\eta\right)\cr
}
\tag{1.1}
$$
where $\Phi_{A}$ depends only on the
spins and disorder variables in $A$.
We assume for simplicity {\bf finite range}, i.e. 
that $\Phi_{A}=0$ for $\hbox{diam}A>r$.
A lot of disordered models can be cast into this form. 

For fixed realization 
of the disorder variable $\eta$ we 
denote by $\mu_{\L}^{\s^{\hbox{\srm b.c.}}}[\eta]$
the corresponding {\bf finite volume Gibbs-measures} in $\L\sb \Z^d$
with boundary condition $\s^{\hbox{\srm b.c.}}$. As usual, they
are the probability measures on $\O$ that are given by the formula 
$$
\eqalign{
&\mu_{\L}^{\s^{\hbox{\srm b.c.}}}[\eta](f)
:=\frac{\sum_{\s_{\L}}f(\s_{\L}\s^{\hbox{\srm b.c.}}_{\Z^d\ba \L})
e^{-\sum_{A\cap \L\neq \em}
\Phi_{A}( \s_{\L}\s^{\hbox{\srm b.c.}}_{\Z^d\ba \L},
\eta)
}
}{
\sum_{\s_{\L}}
e^{-\sum_{A\cap \L\neq \em}
\Phi_{A}( \s_{\L}\s^{\hbox{\srm b.c.}}_{\Z^d\ba \L},
\eta)
}
} 
}
\tag{1.2}
$$
for any bounded measurable observable $f:\O\rightarrow \R$.
The finite-volume 
summation is over $\s_{\L}\in \O_{0}^{\L}$. 
The symbol $\s_{\L}\s^{\hbox{\srm b.c.}}_{\Z^d\ba \L}$
denotes the configuration in $\O$ that is given 
by $\s_{x}$ for $x\in\L$ and by
$\s^{\hbox{\srm b.c.}}_{x}$ for $x\in \Z^d\ba \L$.

We look at spins and disorder variables at the same
time and define {\bf joint spin variables}
$\xi_x=(\s_x,\eta_x)\in \O_{0}\times \HH_{0}$. 
The objects of main interest will then be  
the corresponding {\bf finite volume joint measures} 
$K_{\L}^{\s^{\hbox{\srm b.c.}}}$. They are the probability
measures on the skew space $\left(\O_{0}\times \HH_{0} \right)^{\Z^d}$
that are given by the formula
$$
\eqalign{
&K_{\L}^{\s^{\hbox{\srm b.c.}}}(F)
:=\int\P(d\eta) \int\mu_{\L}^{\s^{\hbox{\srm b.c.}}}[\eta](d\s)
F(\s,\eta) 
}
\tag{1.3}
$$
for any bounded measurable joint observable  
$F:\O\times\HH\rightarrow \R$.
We will consider the following
examples in more detail:

\medskip

\item{\bf (i)} {\bf The Random-Field Ising Model:} 
The single spin space 
is $\O_0=\{-1,1\}$.
The Hamiltonian is 
$$
\eqalign{
&H^{\eta}\left(\s\right)
=-J\sum_{<x,y>}\s_x \s_y -h\sum_{x}\eta_x \s_x
}
\tag{1.4}
$$
where the formal sum is over nearest neighbors $<x,y>$ and $J,h>0$. 
The disorder variables are given by the random fields $\eta_x$ 
that are i.i.d. with single-site distribution $\nu$
that is supported on a finite set $\HH_0$.
  



\item{} The joint spins we will consider are  
given in a natural way 
by the Ising spin and the random field
at the same site, i.e. $\xi_x=(\s_x,\eta_x)$.
$\xi_x$ is thus $4$-valued in the case of symmetric Bernoulli distribution.  

\bigskip

\item{\bf (ii)} {\bf Ising Models with Random Couplings: Random Bond, EA-Spinglass} 

The single spin space is $\O_0=\{-1,1\}$.
The Hamiltonian is 
$$
\eqalign{
&H^{\eta}\left(\s\right)
=-\sum_{x,e}J_{x,e}\s_x \s_{x+e}
}
\tag{1.5}
$$
where the formal sum is over sites $x\in \Z^d$ and 
the nearest neighbor vectors in the positive
lattice directions, i.e. $e\in\{(1,0,0,\dots,0),
(0,1,0,\dots,0),\dots,(0,0,\dots,1)\}=:\EE$. 
The random variables
$J_{x,e}$ take finitely many values, independently
over the `bonds' $x,e$.
Specific distributions we will consider are e.g.

\item{(a)} Random Bond: $J_{x,e}$ takes values $J^1,J^2>0$ 


\item{(b)} EA-Spinglass: Symmetric
(non-degenerate) $3$-valued, 
$J_{x,e}$ takes values $-J,0,J$ with $\nu(J_{x,e}=J)=
\nu(J_{x,e}=-J)$,
$0<\nu(J_{x,e}=0)<1$

\item{} We define the {\bf joint spins} 
by the Ising spin and the collection of adjacent
couplings pointing in the positive direction, i.e. 
$\xi_x=(\s_x,\eta_x)=(\s_x,(J_{x,e})_{e\in \EE})$.
It is thus  $16$-valued in dimension $3$ in case (a).

\bigskip
We think of the Random Field Ising model
for a moment to motivate what we are going to do. 
Recall that, in two dimensions, for almost every
realization of the random fields $\eta$ w.r.t. to the $\P$
there exists a unique infinite volume Gibbs measure $\mu(\eta)$
(see [AW]). In three or more dimensions, for low temperatures
and `small disorder' there exist ferromagnetically
ordered phases $\mu^{+,-}(\eta)$ obtained by different boundary
conditions [BK]. 
Different from the GriSing example of [EMSS] we can hence consider 
various infinite volume versions of the form 
`$\P(d\eta)\mu(\eta)(d\s)$'.

The most general thing now that 
we can reasonably do, 
is to fix {\it any} boundary condition 
$\s^{\hbox{\srm b.c.}}$.
Then, due to compactness, there are always subsequences 
such that the corresponding $K^{\s^{\hbox{\srm b.c.}}}_{\L}(d\x)$ 
converges weakly to a probability measure
on the skew space that we call
$K(d\x)$. 
Note that this measure can in general depend
on the boundary condition and the particular choice of the subsequence in
$d\geq 2$.
It can be shown that: by conditioning  
$K(d\x)=K(d\s,d\eta)$ on the disorder variable 
$\eta$ one obtains a  (not necessarily extremal) random
infinite volume Gibbs-measure, 
for $\P$-almost every $\eta$.\footnote{$^1$}
{A reader who is familiar 
with meta-states will recognize that this measure $K(d\s|\eta)$
is precisely the barycenter of the (corresponding) 
Aizenman-Wehr meta-state, see e.g. Newman 
[N]. For more general information about meta-states and random symmetry breaking 
see [NS1]-[NS4], [K2]-[K5]}
The aim of this paper is to investigate the question: 

{\bf When are the weak limit points of 
$K_{\L}^{\s^{\hbox{\srm b.c.}}}(d\xi)$
Gibbs-measures on the skew-space?
When are they almost [almost not] Gibbs?}

This investigation is about continuity 
properties of conditional expectations. 
Throughout the paper we will use the following notion 
of continuity that involves only uniquely defined finite volume events. 
Following [MRM] we say: 

{\bf Definition: }{\it A point $\xi\in \O\times \HH$ 
is called {\bf good configuration} 
for $K$, if 
$$
\eqalign{
&\sup_{{\xi^+,\xi^-}\atop{\L:\L\sp V}}
\Bigl| 
K(\tilde\xi_{x}\bigl|\xi_{V\ba x},\xi^{+}_{\L\ba V})
-K(\tilde
\xi_{x}\bigl|\xi_{V\ba x},\xi^{-}_{\L\ba V})\Bigr |\rightarrow 0
}
\tag{1.6}
$$
with $V\uparrow\Z^d$, for any site $x\in\Z^d$,
for any $\tilde\xi_{x}\in\HH_0$.
Call $\xi$ {\bf bad }, if it is not good. 
}

As usual we have written
$\xi_{A}=(\xi_x)_{x\in A}$ (and will also do so for 
$\s_{A}$, $\eta_{A}$).

In words: Good configuration are the points $\xi$
where: The family of conditional expectations
of $K$ is equicontinuous w.r.t. the parameter $\L$. 

We recall: If there are no bad configurations, 
the measure $K$ is Gibbsian (see [MRM]). 
If Gibbsianness does not hold, one can 
ask for the $K$-measure of the 
set of bad configurations.  
We say that $K$ is almost Gibbsian, if it has $K$-measure zero.
If it has $K$-measure one, we say that $K$ is almost non-Gibbsian.
(See also the beginning of the next chapter.)

In the remainder of the paper we will prove 
criteria that ensure
that a configuration $(\eta,\s)$ is good or bad (see propositions 1-6).
It might not be very intuitive at first sight
to understand why such measures can ever be non-Gibbsian. 
Let us stress the following facts:
Surely, the conditional expectation of the spin-variable 
$\s_x$ given the joint variable $\x=(\s,\eta)$ 
away from $x$ {\it and} $\eta_x$ is a 
local function, given by the local specifications.   
Trivially, the conditional expectation 
of the disorder variable
$\eta_x$ given $\eta$ away from $x$ is a local function - 
it is even independent. 
However: The conditional expectation of $\eta_x$ given $\eta$ {\it and} $\s$ 
away from $x$ can be highly nontrivial, due to the coupling between 
spins and disorder
arising from the local specifications (1.2).

Rather than presenting our general results at this point, 
we specialize to the Random Field Ising Model.
For this model there is a complete characterization of a bad 
configuration in terms of the behavior 
of the finite volume Gibbs-measures that is particularly
transparent.  We obtain:

\theo{1}{\it Consider a random 
field Ising model of the form (1.4), in any dimension $d$. 
A configuration $\x=(\eta,\s)$
is a bad configuration for {\bf any} joint measure
obtained as a limit point of the finite volume 
joint measures $\P(d \eta)
\mu^{\s^{\hbox{\srm b.c.}}_{\del\L}}_{\L}[\eta]$ 
if and only if 
$$
\eqalign{
\lim_{\L\uparrow\infty}\mu^{+}_{\L}
[\eta_{\L}]\left(\tilde \s_x=1\right)
>\lim_{\L\uparrow\infty}\mu^{-}_{\L}
[\eta_{\L}]\left(\tilde \s_x=1\right)
}
\tag{1.7}
$$
for some site $x$, independent of $\s$.
Here $\mu^{+,-}_{\L}$ are the finite volume Gibbs measures
with $+$ (resp. $-$) boundary conditions.   
}

Note, that the theorem will hold for the joint measures
corresponding to Dobrushin 
states that are supposed to exist in $d\geq 4$.\footnote{$^1$}
{For an existence result of this model
in the SOS-approximation, see [BoK1], [K1]}
Using the known results 
about the random field model one immediately obtains:

\corollary{}{\it

\item{(i)} $d=1$: $K$ is Gibbsian, for all $J$, $h>0$.

\item{(ii)} $d=2$: $K$ is a.s. Gibbsian for all $J$, $h>0$.

\item{} On the other hand, suppose that $\nu[\eta_x=0]>0$.
Assume that $J$ is sufficiently large and $h>0$. 
Then $K$ is not Gibbsian. 

\item{(iii)} $d\geq 3$, $\nu$ symmetric, 
$J>0$ sufficiently large, $\nu[\eta^2_x]$
sufficiently small. 
Then any such $K$ is a.s. not Gibbs.

}

Indeed: The a.s. Gibbsianness in $d=2$ follows from the a.s. absence 
of ferromagnetism, proved in [AW]. 
That we have 
Non-Gibbsianness in $d\geq 2$ if the support 
of the random fields contains zero follows from
the fact that the configuration $\x=(\eta_x\equiv 0,\s)$ is 
a bad, if $J$ is large enough s.t. there
is ferromagnetic order in the homogeneous Ising ferromagnet.   
A.s. non-Gibbsianness under the conditions (iii)
follows from the existence ferromagnetic order, proved in [BK].

The organization of the paper is as follows.
In Chapter II we investigate the one-site conditional
probabilities of $K$ and prove general criteria 
that ensure that a configuration is good or bad. 
We will see that the important general step is to 
consider the single-site variation of the Hamiltonian
w.r.t. the disorder variable $\eta_x$ and rewrite
the conditional expectations in the form of Lemma 1. 
This leads to expressions involving certain expectations 
of the `conjugate' spin-observable. 
In the example of the random field model this observable is
just the spin $\s_x$; thus the corresponding criteria
in Theorem (i)
are simply formulated in terms of the magnetization.

In Chapter III we apply our results. We prove Theorem 1 about 
the RFIM.
Next we comment on Models with decoupling configurations,
recalling the GriSing random field of [EMSS] and 
Models with random couplings (including spinglasses)
that can be zero. This provides more examples
of non-Gibbsian fields. 
Next we specialize our criteria of Chapter II 
to Models with random couplings, proving Theorem 2. 
Based on this we give a heuristic discussion explaining
how the validity of the Gibbsian property can be linked
to the absence of random Dobrushin states.

\bigskip

\chap{Acknowledgments:}
The author thanks A.van Enter 
for a private explanation of reference [EMSS].

\bigskip
\bigskip


\bigskip\bigskip

\chap{II. Criteria for joint [non-]Gibbsianness \hfill}

In this chapter we are going to investigate
whether a configuration $\x=(\eta,\s)$ is good 
or bad for the joint states $K$. 
We will obtain criteria that 
are given in terms of the local specifications. 
To do so we introduce the single-site variation 
of the Hamiltonian w.r.t. disorder (2.2) and use 
the finite volume perturbation formula (2.3)
to rewrite the conditional expectations of $K$ in 
the form of Lemma 1. This leads to the 
characterization of good resp. bad configurations 
of the Corollary of Proposition 1. 
As direct consequences thereof, Propositions 2 and 3 
give more convenient 
conditions that ensure goodness resp. badness. 
Under the additional assumption 
of a.s. convergent Gibbs measures
we obtain the slightly less obvious 
criterion for badness of Proposition 4.




Before we start, 
let us however summarize the following facts
about the notion of good configuration
and its relevance for Gibbsianness, for the sake of 
clarity: 

\item{(i)} If $\x$ is {\it bad} for $K$
any version
of the conditional expectation $\xi_{\Z^d}
\mapsto K(\xi_x|\xi_{\Z^d\ba x})$
must be  discontinuous for some site $x$
(use DLR-equation, see Proposition 4.3[MRM]). 

\item{(ii)} Conversely: 
Assume that $\hat\x\in \GG:=\{\xi;\xi\hbox{ is good}\}$.
Then $\lim_{\L\uparrow\Z^d}K(\xi_x|\hat\xi_{\L\ba x})$
exists for any site $x$ and hence also 
$\lim_{\L\uparrow\Z^d}K(\xi_{V}|\hat\xi_{\L\ba V})=:
\g_{V}(\xi_{V}|\hat\xi_{\Z^d\ba V})$
exists for any finite volume $V$.
If $\GG$ has full measure w.r.t $K$,
the above limit can be (arbitrarily) extended to 
a measurable function of the conditioning.
It is readily seen to define a version of the conditional 
expectation $\xi_{\Z^d\ba V}\mapsto K(\xi_{V}|\xi_{\Z^d\ba V})$
that is continuous within the set 
$\GG$ 
[i.e.: $\xi^{(N)}\rightarrow \xi$ with $\xi^{(N)},\xi\in \GG$
implies $K(\xi_{V}|\xi_{\Z^d\ba V}^{(N)})\rightarrow 
K(\xi_{V}|\xi_{\Z^d\ba V})$]. (See [MRM]: Proof of 
Proposition 4.4).
In this situation we call $K$ {\bf almost Gibbs}.
\footnote{$^1$}{If $K(\GG)=1$ but $\GG\neq \HH\times \O$,
we have:
$\GG$ is dense in $\HH\times \O$
[since any ball w.r.t. a metric for the product
topology has to have 
positive $K$-mass, under the assumption of bounded interactions $\Phi$.]
Thus the conditional expectation is continuous on $\GG$ 
but necessarily {\it not} uniformly continuous (because it could
be extended to the whole space otherwise.)}

\item{} In particular: If {\bf every} configuration is good,  
the measure $K$ has a version
of the conditional expectation that is
continuous on the whole space and is {\bf Gibbs} therefor.   

\bigskip

\noindent In the sequel it will be important to keep
track of the local dependence of various quantities. 
It will be useful to make this explicit. We use the following

{\bf Notation:} 
For the fixed interaction range $r$ we introduce the $r$-boundary 
$\del B=\{x\in \Z^d\ba B; d(x,B)\leq r\}$.
In the same fashion we write
$\ov{B}=B\cup \del B$ and 
$\del_{-} B=\{x\in B; d(x,B^c)\leq r\}$, 
$B^{o}= B\ba \del_{-} B$.

In this way we will write e.g. 
$K_{\L}^{\s^{\hbox{\srm b.c.}}}(\s_{\L},\eta_{\ov\L})
=\P_{\ov\L}(\eta_{\ov\L})
\mu_{\L}^{\s^{\hbox{\srm b.c.}}_{\del\L}}[\ov\eta_{\L}](\s_{\L})$
to denote the corresponding probabilities.

To investigate the quantity (1.6) for 
the infinite volume joint measure we 
will look at 
$K_{\L_N}^{\s^{\hbox{\srm b.c.}}_{\del\L_N}}$ 
with finite $\L_N$. 
Next, to investigate the conditional distributions of $\xi_x$
it suffices to look at the conditional distributions
of $\eta_x$. Indeed, we may write (for sufficiently large $\L_N$) 
$$
\eqalign{
&K_{\L_N}^{\s^{\hbox{\srm b.c.}}_{\del\L_N}}\left[\s_x;\eta_{x}\bigl|
\s_{\L\ba x};\eta_{\L\ba x}\right]
=K_{\L_N}^{\s^{\hbox{\srm b.c.}}_{\del\L_N}}
\left[\s_x\bigl|\s_{\L\ba x};\eta_x,\eta_{\L\ba x}\right]
\times 
K_{\L_N}^{\s^{\hbox{\srm b.c.}}_{\del\L_N}}
\left[\eta_x\bigl|\s_{\L\ba x};\eta_{\L\ba x}\right]\text{where}\cr
&K_{\L_N}^{\s^{\hbox{\srm b.c.}}_{\del\L_N}}
\left[\s_x\bigl|\s_{\L\ba x};\eta_x,\eta_{\L\ba x}\right]\cr
&=\frac{\E_{\ov{\L_N}\ba \L}\mu^{\s^{\hbox{\srm b.c.}}_{\del\L_N}}_{\L_N}
[\eta_x,\eta_{\L\ba x},\tilde\eta_{\ov{\L_N}\ba \L}](\s_x,\s_{\L\ba x})}
{\sum_{\s'_x}\E_{\ov{\L_N}\ba \L}\mu^{\s^{\hbox{\srm b.c.}}_{\del\L_N}}_{\L_N}
[\eta_x,\eta_{\L\ba x},\tilde\eta_{\ov{\L_N}\ba \L}](\s'_x,\s_{\L\ba x})}
=\mu^{\s_{\del x}}_{x}[\eta_{x},\eta_{\del {x}}](\s_{x})\cr
}
\tag{2.1}
$$
where the second equality follows 
from the application of the compatibility relation
for the $\mu$-measures for the inner volume made of
the single site $x$,
as soon  as $\L\sp \ov{x}$. 
There is of course no non-locality as a function 
of $\s_{\L\ba x},\eta_{\L\ba x}$ in this term. 

On the other hand we see that, if the conditional 
$\eta_x$-distribution
has a non-local behavior as a function 
of $\s_{\L\ba x},\eta_{\L\ba x}$, this carries 
over also to the $\s_x$-marginal
$K_{\L_N}^{\s^{\hbox{\srm b.c.}}_{\del\L_N}}
\left[\s_x\bigl|\s_{\L\ba x};\eta_{\L\ba x}\right]
=\int K_{\L_N}^{\s^{\hbox{\srm b.c.}}_{\del\L_N}}
\left[d\tilde\eta_x\bigl|\s_{\L\ba x};\eta_{\L\ba x}\right]
\mu^{\s_{\del x}}_{x}[\tilde\eta_{x},\eta_{\del {x}}](\s_{x})
$
unless the dependence on $\tilde \eta_x$ of the one-site
expectation under the last integral is trivial, of course.

After these simple remarks 
we come to the important formula that is going to be 
the starting point of all our analysis. 

Let us define the {\it single-site-variation of the Hamiltonian w.r.t. 
the disorder variable}\footnote{$^1$}
{A quantity of this type also plays a crucial role
in [AW] where the fluctuations 
of extensive quantities are investigated. 
Its Gibbs expectation could be termed
`order parameter that is conjugate to the disorder'.} 
$\eta_x$ at the site $x$ to be
$$
\eqalign{
&\D H_{x}(\s_{\ov{x}},\eta_{x},\eta_{x}^{0},\eta_{\del {x}})
= \sum_{A; A\ni x}\Bigl[\Phi_{A}\left(
\s_{\ov{x}},\eta_{x}\eta_{\del {x}}\right) 
-\Phi_{A}\left(
\s_{\ov{x}},\eta_{x}^{0}\eta_{\del {x}}
\right)
\Bigr]
\cr
}
\tag{2.2}
$$
where  is some fixed 
reference configuration (that is independent of $x$). 
While we will later put $\eta_x^0\in \HH^0$
one might also want to choose some other value 
that is not in the support of the single-site distribution
in certain situations.

The trick is to use the `finite volume
perturbation formula'  
$$
\eqalign{
&\int\mu^{\s^{\hbox{\srm b.c.}}_{\del\L}}_{\L}
[\eta_x,\eta_{\ov{\L}\ba x}]
(d\s_{\L})f(\s_{\L})
=\frac{
\int\mu^{\s^{\hbox{\srm b.c.}}_{\del\L}}_{\L}
[\eta_{x}^{0},\eta_{\ov{\L}\ba x}]
(d\s_{\L})f(\s_{\L})e^{
-\D H_{x}(\s_{\ov{x}},\eta_{x},\eta_{x}^{0},\eta_{\del {x}})
}
}{
\int\mu^{\s^{\hbox{\srm b.c.}}_{\del\L}}_{\L}
[\eta_{x}^{0},\eta_{\ov{\L}\ba x}]
(d\s_{\L})e^{
-\D H_{x}(\s_{\ov{x}},\eta_{x},\eta_{x}^{0},\eta_{\del {x}})
}
}
\cr
}
\tag{2.3}
$$
which is just a rewriting of Boltzmann factors. 
Using this we get
\medskip

\lemma{1}{For any reference configuration $\eta_{x}^{0}$
the conditional expectations of the one-site
disorder variable $\eta_x$ can be rewritten as
$$
\eqalign{
&K_{\L_N}^{\s^{\hbox{\srm b.c.}}_{\del\L_N}}
\left[\eta_{x}\bigl|\s_{\L\ba x};\eta_{\L\ba x}\right]\cr
&=\nu(\eta_x)\int\mu^{\s_{\del x}}_{x}
[\eta_{x}^{0},\eta_{\del x}](d\tilde\s_{x})e^{
-\D H_{x}(\s_{\del{x}},\tilde \s_{x},\eta_{x},\eta_{x}^{0},\eta_{\del {x}})
}
\cr
&\qquad\times\int K_{\L_N}^{\s^{\hbox{\srm b.c.}}_{\del\L_N}}
\left[d\tilde\eta_{\ov{\L_N}\ba\L}\bigl|\s_{\del_{-}\L};\eta_{x}^{0},\eta_{\L\ba x}
\right]
\left[\int\mu^{\s^{\hbox{\srm b.c.}}_{\del\L_N}}_{\L_N}
[\eta_{x}^{0},\eta_{\L\ba x},\tilde\eta_{\ov{\L_N}\ba \L}](d\tilde\s_{\ov{x}})
e^{-\D H_{x}(\tilde\s_{\ov{x}},\eta_{x},\eta_{x}^{0},\eta_{\del {x}})}
\right]^{-1}\cr
&\times\Biggl\{\sum_{\eta'_x}\nu(\eta'_x)
\int\mu^{\s_{\del x}}_{x}
[\eta_{x}^{0},\eta_{\del x}](d\tilde\s_{x})e^{
-\D H_{x}(\s_{\del{x}},\tilde \s_{x},\eta'_{x},\eta_{x}^{0},\eta_{\del {x}})
}
\cr
&\qquad\times\int K_{\L_N}^{\s^{\hbox{\srm b.c.}}_{\del\L_N}}
\left[d\tilde\eta_{\ov{\L_N}\ba\L}\bigl|\s_{\del_{-}\L};\eta_{x}^{0},\eta_{\L\ba x}
\right]
\left[\int\mu^{\s^{\hbox{\srm b.c.}}_{\del\L_N}}_{\L_N}
[\eta_{x}^{0},\eta_{\L\ba x},\tilde\eta_{\ov{\L_N}\ba \L}](d\tilde\s_{\ov{x}})
e^{-\D H_{x}(\tilde\s_{\ov{x}},\eta'_{x},\eta_{x}^{0},\eta_{\del {x}})}
\right]^{-1}
\Biggr\}^{-1}
}
\tag{2.4}
$$
}
\proof To compute the conditional distribution of $\eta_x$
we use the finite volume perturbation formula 
to extract the variation of $\eta_x$. We use 
a convention to put tildes on quantities that are integrated
and write 
$$
\eqalign{
&K_{\L_N}^{\s^{\hbox{\srm b.c.}}_{\del\L_N}}
\left[\s_{\L\ba x};\eta_x,\eta_{\L\ba x}\right]
=\P(\eta_x)\P(\eta_{\L\ba x})\times
\E_{\ov{\L_N}\ba \L}\mu^{\s^{\hbox{\srm b.c.}}_{\del\L_N}}_{\L_N}
[\eta_x,\eta_{\L\ba x},\tilde\eta_{\ov{\L_N}\ba \L}](\s_{\L\ba x})\cr
&=\P(\eta_x)\P(\eta_{\L\ba x})\times\E_{\ov{\L_N}\ba \L}
\frac{\int\mu^{\s^{\hbox{\srm b.c.}}_{\del\L_N}}_{\L_N}
[\eta_{x}^{0},\eta_{\L\ba x},\tilde\eta_{\ov{\L_N}\ba \L}]
(d\tilde\s_\L)e^{-\D H_{x}(\tilde\s_{\ov{x}},\eta_{x},\eta_{x}^{0},\eta_{\del {x}})}1_{\tilde \s_{\L\ba x}=\s_{\L\ba x}}
}
{
\int\mu^{\s^{\hbox{\srm b.c.}}_{\del\L_N}}_{\L_N}
[\eta_{x}^{0},\eta_{\L\ba x},\tilde\eta_{\ov{\L_N}\ba \L}](d\tilde\s_{\L})
e^{-\D H_{x}(\tilde\s_{\ov{x}},\eta_{x},\eta_{x}^{0},\eta_{\del {x}})}
}
\cr
&=\P(\eta_x)\times\P(\eta_{\L\ba x})\mu^{\s_{\del_{-}\L}}_{\L^{o}}
[\eta_{x}^{0},\eta_{\L\ba x}](\s_{\L^o\ba x})\cr
&\quad\times \int\mu^{\s_{\del x}}_{x}
[\eta_{x}^{0},\eta_{\del x}](d\tilde\s_{x})e^{
-\D H_{x}(\s_{\del{x}},\tilde \s_{x},\eta_{x},\eta_{x}^{0},\eta_{\del {x}})
}\cr
&\times\E_{\ov{\L_N}\ba \L}
\frac{\mu^{\s^{\hbox{\srm b.c.}}_{\del\L_N}}_{\L_N}
[\eta_{x}^{0},\eta_{\L\ba x},\tilde\eta_{\ov{\L_N}\ba \L}](\s_{\del_{-}\L})
}
{\int\mu^{\s^{\hbox{\srm b.c.}}_{\del\L_N}}_{\L_N}
[\eta_{x}^{0},\eta_{\L\ba x},\tilde\eta_{\ov{\L_N}\ba \L}](d\tilde\s_{\L})
e^{-\D H_{x}(\tilde\s_{\ov{x}},\eta_{x},\eta_{x}^{0},\eta_{\del {x}})}}
\cr
}
\tag{2.5}
$$
We have used the compatibility relations for the 
local specifications in the last equation
and we have assumed that $\L,\L_N$ are sufficiently large.
To get the conditional 
expectation we need to normalize the r.h.s. by its $\eta_x$-sum.
To see that the claim follows now note that 
$$
\eqalign{
&\E_{\ov{\L_N}\ba \L}
\frac{\mu^{\s^{\hbox{\srm b.c.}}_{\del\L_N}}_{\L_N}
[\eta_{x}^{0},\eta_{\L\ba x},\tilde\eta_{\ov{\L_N}\ba \L}](\s_{\del_{-}\L})
}
{\int\mu^{\s^{\hbox{\srm b.c.}}_{\del\L_N}}_{\L_N}
[\eta_{x}^{0},\eta_{\L\ba x},\tilde\eta_{\ov{\L_N}\ba \L}](d\tilde\s_{\L})
e^{-\D H_{x}(\tilde\s_{\ov{x}},\eta_{x},\eta_{x}^{0},\eta_{\del {x}})}}\cr
&= 
\int K_{\L_N}^{\s^{\hbox{\srm b.c.}}_{\del\L_N}}
\left[d\tilde\eta_{\ov{\L_N}\ba\L}\bigl|\s_{\del_{-}\L};\eta_{x}^{0},\eta_{\L\ba x}
\right]
\left[\int\mu^{\s^{\hbox{\srm b.c.}}_{\del\L_N}}_{\L_N}
[\eta_{x}^{0},\eta_{\L\ba x},\tilde\eta_{\ov{\L_N}\ba \L}](d\tilde\s_{\ov{x}})
e^{-\D H_{x}(\tilde\s_{\ov{x}},\eta_{x},\eta_{x}^{0},\eta_{\del {x}})}
\right]^{-1}\cr
&\quad\times 
\E_{\ov{\L_N}\ba \L}
\mu^{\s^{\hbox{\srm b.c.}}_{\del\L_N}}_{\L_N}
[\eta_{x}^{0},\eta_{\L\ba x},\tilde\eta_{\ov{\L_N}\ba \L}](\s_{\del_{-}\L})
}
\tag{2.6}
$$
where the term in the last line is 
just a constant for $\eta_x$.
\endproof

\bigskip
{\bf Remark: }
The formula gives the modification 
of the conditional expectation compared
with the `free' a-priori measure $\nu(\eta_x)$
that results from the non-trivial coupling 
of $\eta$ to the spin-variable $\s$.
The second term in the second line of (2.4),
a Gibbs expectation of the exponential 
of the single-site variation
of the Hamiltonian, is of course a local function in the
conditioning. Assuming the finiteness of the potential
it is bounded. Thus, to investigate the potential
non-locality of the l.h.s. one has to investigate
the third line of (2.4).
\medskip
{\bf Remark: } The {\it local} $\L_N$-limit of the 
conditional expectation 
$K_{\L_N}^{\s^{\hbox{\srm b.c.}}_{\del\L_N}}
\left[d\tilde\eta_{\ov{\L_N}\ba\L}\bigl|\s_{\del_{-}\L};\eta_{x}^{0},\eta_{\L\ba x}
\right]$ exists from 
the assumption of the existence of the joint local 
$\lim_{\L_N\uparrow\Z^d}K_{\L_N}^{\s^{\hbox{\srm b.c.}}_{\del\L_N}}$.
Also, the $\L_N$-limit of the complete third line
of (2.4) [that involves the average of an $N$-dependent
function of $\tilde\eta$] exists: 
The $\L_N$-limit of the quantity
in the last line of (2.5) exists by our assumption
on the existence of a $\L_N$-limit on the l.h.s. of (2.5).
The $\L_N$ limit of the last line of (2.6) [the normalization needed 
to obtain probabilities] 
also exists by the hypothesis. 

\bigskip

Sometimes it is convenient to rewrite (2.4) using that, 
by the finite volume perturbation formula, we have 
$$
\eqalign{
&\left[\int\mu^{\s^{\hbox{\srm b.c.}}_{\del\L_N}}_{\L_N}
[\eta_{x}^{0},\eta_{\L\ba x},\tilde\eta_{\ov{\L_N}\ba \L}](d\tilde\s_{\ov{x}})
e^{-\D H_{x}(\tilde\s_{\ov{x}},\eta_{x},\eta_{x}^{0},\eta_{\del {x}})}
\right]^{-1}\cr
&=\int\mu^{\s^{\hbox{\srm b.c.}}_{\del\L_N}}_{\L_N}
[\eta_x,\eta_{\L\ba x},\tilde\eta_{\ov{\L_N}\ba \L}](d\tilde\s_{\ov{x}})
e^{+\D H_{x}(\tilde\s_{\ov{x}},\eta_{x},\eta_{x}^{0},\eta_{\del {x}})}
\equiv \mu^{\s^{\hbox{\srm b.c.}}_{\del\L_N}}_{\L_N}
[\eta_{\L},\tilde\eta_{\ov{\L_N}\ba \L}]\left(
e^{\D H_{x}(\eta_{x},\eta_{x}^{0},\eta_{\del {x}})}\right)
}
\tag{2.7}
$$
The reader may also want to note that (2.7) is just 
a fraction of two partition functions, 
$Z^{\s^{\hbox{\srm b.c.}}_{\del\L_N}}_{\L_N}
[\eta^0_x\eta_{\L\ba x}
\tilde\eta_{\ov{\L_N}\ba \L}]      
/Z^{\s^{\hbox{\srm b.c.}}_{\del\L_N}}_{\L_N}
[\eta_x\eta_{\L\ba x}
\tilde\eta_{\ov{\L_N}\ba \L}]$  (using usual notations) 
which makes the symmetry between $\eta_x$ and $\eta_x^{0}$
more apparent.

\medskip
From this we have

\proposition{1}{
$$
\eqalign{
&\frac{K\left[\eta_{x}^1\bigl|\s_{\L\ba x};\eta_{\L\ba x}\right]}
{K\left[\eta^2_{x}\bigl|\s_{\L\ba x};\eta_{\L\ba x}\right]}
=q^{\hbox{\srm local}}(\eta^1_x,\eta^2_x,\s_{\del{x}},\eta_{\del {x}} )
\,\,q^{\hbox{\srm nonloc} }_{\L,x}[\eta^1_x,\eta^2_x,
\eta_{\L\ba x},\s_{\del_{-}\L}]
\cr
}
\tag{2.8}
$$
where
$$
\eqalign{
&q^{\hbox{\srm local}}(\eta^1_x,\eta^2_x,\s_{\del{x}},\eta_{\del {x}})
=\frac{\nu(\eta^1_x)}{\nu(\eta^2_x)}
\int\mu^{\s_{\del x}}_{x}
[\eta^2_{x},\eta_{\del x}](d\tilde\s_{x})e^{
-\D H_{x}(\s_{\del{x}},\tilde \s_{x},\eta^1_{x},\eta^2_{x},\eta_{\del {x}})
}
}
\tag{2.9}
$$
is a local function of $\s,\eta$ and
$$
\eqalign{
&q^{\hbox{\srm nonloc} }_{\L,x}[\eta^1_x,\eta^2_x,\eta_{\L\ba x},\s_{\del_{-}\L}]\cr
&=\lim_{\L_N\uparrow\Z^d}
\int K_{\L_N}^{\s^{\hbox{\srm b.c.}}_{\del\L_N}}
\left[d\tilde\eta_{\ov{\L_N}\ba\L}\bigl|\s_{\del_{-}\L};\eta^2_{x},\eta_{\L\ba x}
\right]
\int\mu^{\s^{\hbox{\srm b.c.}}_{\del\L_N}}_{\L_N}
[\eta^1_{x},\eta_{\L\ba x},\tilde\eta_{\ov{\L_N}\ba \L}](d\tilde\s_{\ov{x}})
e^{\D H_{x}(\tilde\s_{\ov{x}},\eta^1_{x},\eta^2_{x},\eta_{\del {x}})}\cr
}
\tag{2.10}
$$
is a potentially nonlocal function of $\s,\eta$.
The last limit exists. 
}

\corollary{\it A point $\xi=(\s,\eta)$ is a good configuration 
for $K$ if and only if
$$
\eqalign{
&\sup_{{\eta^+,\eta^-;\s^+,\s^- }
\atop{\L:\L\sp V}}
\Biggl|q^{\hbox{\srm nonloc} }_{\L,x}[\eta^{1}_x,\eta^{2}_x,
\eta_{V\ba x},
\eta^+_{\L\ba V},\s^+_{\del_{-}\L}]
-q^{\hbox{\srm nonloc} }_{\L,x}[\eta^{1}_x,\eta^{2}_x,
\eta_{V\ba x},
\eta^-_{\L\ba V},\s^-_{\del_{-}\L}]
\Biggr|\rightarrow 0
}
\tag{2.11}
$$
with $V\uparrow\Z^d$, for any site $x\in\Z^d$,
for any pair $\eta^{1}_{x},\eta^{2}_{x}\in\HH_0$.
}

\proof To prove the proposition choose the reference configuration
$\eta_x^{0}=\eta^2_x$ and use Lemma 1, along with (2.7). 
The Corollary follows
from the fact that $q^{\hbox{\srm local}}$
is a local function, and that it suffices to check
the conditional expectations of the disorder variable by
(2.1). Note to this end that both $q$'s
in Proposition 0 
are uniformly bounded against zero and one, by 
the assumed finiteness of $\D H_x$.
\endproof  

To understand the symmetry between $\eta^1$ and $\eta^2$ 
in this formula we remark that 
$q^{\srm local}$ as well as the inner integral in (2.10)
can be written as fractions of partitions functions, 
by the remark following (2.7). 
We will now 
discuss various consequences of Corollary of Proposition 1.
It is very difficult to say anything 
reasonable about the behavior 
of the conditional measure 
$K_{\L_N}^{\s^{\hbox{\srm b.c.}}_{\del\L_N}}
\left[d\tilde\eta_{\ov{\L_N}\ba\L}\bigl|\s_{\del_{-}\L};\eta^2_{x},\eta_{\L\ba x}
\right]$, as a function of the spin-conditioning
$\s_{\del_{-}\L}$.  
So, in our examples we will at first draw conclusions
from estimates that are {\it uniform} w.r.t. 
the integration variable $\tilde\eta_{\L_N\ba\L}$. 

We start with a
criterion for points $\xi=(\eta,\s)$ being
good configurations that is a pretty much
straightforward consequence of Proposition 1. 
This will be employed if we want to show Gibbsianness. 
Below will give a slightly more complicated
criterion for points $\xi=(\eta,\s)$ being
bad configurations, needed to investigate 
non-Gibbsianness. 

\bigskip

\proposition{2}{\it 
Suppose that $\eta$ is such that, for any $x\in\Z^d$, we have that 
$$
\eqalign{
&r_{V,x}(\eta^1_{x},\eta^2_{x},\eta)
:=\sup_{{\eta^+,\eta^-}\atop{\L:\L\sp V}}
\Biggl|\int\mu^{\s^{\hbox{\srm b.c.}}_{\del\L}}_{\L}
[\eta^1_{x},\eta_{V\ba x},\eta^+_{\ov{\L}\ba V}]
\left(
e^{\D H_{x}(\eta^1_{x},\eta^2_{x},\eta_{\del {x}})}
\right)\cr
&-\int\mu^{\s^{\hbox{\srm b.c.}}_{\del\L}}_{\L}
[\eta^1_{x},\eta_{V\ba x},\eta^-_{\ov{\L}\ba V}]
\left(
e^{\D H_{x}(\eta^1_{x},\eta^2_{x},\eta_{\del {x}})}
\right)
\Biggr|\rightarrow 0
}
\tag{2.12}
$$
with $V\uparrow\Z^d$, for any $x$,
for any pair $\eta^{1}_{x},\eta^{2}_{x}\in\HH_0$.
Then the configuration $\eta,\s$
is a good configuration, for any $\s$.
}

\proof To see that the hypothesis implies (2.11)
we use that 
$$
\eqalign{
&
\Biggl|\mu^{\s^{\hbox{\srm b.c.}}_{\del\L_N}}_{\L_N}
[\eta^1_{x},\eta_{V\ba x},
\eta^{+,-}_{\L\ba V},\tilde\eta_{\ov{\L_N}\ba \L}]
\left(e^{\D H_{x}(\eta^1_{x},\eta^2_{x},\eta_{\del {x}})}\right)\cr
&\quad- \mu^{\s^{\hbox{\srm b.c.}}_{\del\L_N}}_{\L_N}
[\eta^1_{x},\eta_{\ov{\L_N}\ba x}]
\left(e^{\D H_{x}(\tilde\s_{\ov{x}},\eta^1_{x},\eta^2_{x},\eta_{\del {x}})}
\right)
\Biggr |
\leq r_{V,x}(\eta^1_{x},\eta^2_{x},\eta)
}
\tag{2.13}
$$
to compare the $\mu$-terms under the $\tilde \eta$-integrals
with a term that is independent of $\tilde\eta$ and $\eta^{+,-}$.
This shows that (2.11)
is bounded by $2r_{V,x}$ which converges 
to zero.
\endproof

\remark To estimate $r_{V,x}(\eta^1_{x},\eta^2_{x},\eta)$ we 
can also bound the variation of the random couplings
by the variation over the boundary conditions 
$$
\eqalign{
&r_{V,x}(\eta^1_{x},\eta^2_{x},\eta)
\leq\sup_{\s^1,\s^2}\left|
\mu^{\s^{1}_{\del_{-} V}}_{V^o}
[\eta^1_x\eta_{V\ba x}]
\left(
e^{\D H_{x}(\eta^1_{x},\eta^2_{x},\eta_{\del {x}})}
\right)
-\mu^{\s^{2}_{\del_{-} V}}_{V^o}
[\eta^1_x\eta_{V\ba x}]
\left(
e^{\D H_{x}(\eta^1_{x},\eta^2_{x},\eta_{\del {x}})}
\right)\right|
}
\tag{2.14}
$$
\bigskip

\remark We see, how (2.12) parallels (1.6).
The quantity that is of interest 
is now the Gibbs-expectation of the 
exponential of the single-site variation as a function
of the disorder variables.  
In words: If we have equicontinuity in the parameter $\L$ of these 
finite $\L$-Gibbs expectations w.r.t. the disorder
variable at the point $\eta$, we conclude
that $\eta,\s$ is a good configuration.  
The reader may also find it intuitive to 
rewrite the Gibbs-expectations appearing in (2.12)
in the form of fractions of partition functions, or (equivalently) 
as exponentials of differences of free energies  taken for 
$\eta_x^1$ and $\eta^2_x$. 
In slightly different words the criterion thus requires:
Equicontinuity in the volume
of the single site-variations of the free energies
w.r.t. the disorder variable at the point $\eta$.


\bigskip 

To get a criterion for bad configurations
that is independent of the behavior of the outer expectation 
of $q^{\srm nonloc}$ [see (2.10)] leads to an expression that is 
slightly more complicated because it contains
an additional supremum. 
\bigskip

\proposition{3}{\it 
Put 
$$
\eqalign{
&q^{\hbox{\srm upper} }_{\L,x}[\eta^1_x,\eta^2_x,\eta_{\L\ba x}]:=
\limsup_{\L_N\uparrow\Z^d}
\sup_{\tilde \eta_{\ov{\L_N}\ba \L}}
\mu^{\s^{\hbox{\srm b.c.}}_{\del\L_N}}_{\L_N}
[\eta^1_x,\eta_{\L\ba x},\tilde \eta_{\ov\L_N\ba \L}]
\left(
e^{\D H_{x}(\eta^1_{x},\eta^2_{x},\eta_{\del {x}})}
\right)
}
\tag{2.15}
$$
Then $\eta,\s$ is a bad configuration for $K$, if for some site $x$, for some
pair $\eta^1_x$, $\eta^2_x$
$$
\eqalign{
&\lim_{V\uparrow\Z^d}
\sup_{{\eta^+,\eta^-}\atop{\L:\L\sp V}}
\left(
\left(q^{\hbox{\srm upper} }_{\L,x}[\eta^2_x,\eta^1_x,\eta_{V\ba x}, \eta^+_{\L\ba V}]
\right)^{-1}
-q^{\hbox{\srm upper} }_{\L,x}[\eta^1_x,\eta^2_x,\eta_{V\ba x}, 
\eta^-_{\L\ba V}]
\right)>0
}
\tag{2.16}
$$
}

\proof By (2.7) and the uniform estimate 
of the $\tilde\eta$-integral we see that 
that 
$$
\eqalign{
&q^{\hbox{\srm nonloc} }_{\L,x}[\eta^1_x,\eta^2_x,\eta_{\L\ba x},\s_{\del_{-}\L}]
\leq q^{\hbox{\srm upper} }_{\L,x}[\eta^1_x,\eta^2_x,\eta_{\L\ba x}],\quad
\geq q^{\hbox{\srm upper} }_{\L,x}[\eta^2_x,\eta^1_x,\eta_{\L\ba x}]^{-1}
}
\tag{2.17}
$$
Hence the claim (discontinuity of the l.h.s.)
follows from the definition of a bad configuration. 
\endproof

\bigskip

\line{\bf 
Models with a.s. convergent Gibbs states:\hfill}

\bigskip

Suppose that we have the existence of a weak limit 
$$
\eqalign{
&\lim_{\L\uparrow\Z^d}
\mu^{\s^{\hbox{\srm b.c.}}_{\del\L}}_{\L}[\eta_{\L}]
= \mu_{\infty}[\eta_{\Z^d}]
}
\tag{2.18}
$$
for $\P$-a.e. $\eta$. It follows that $\mu_{\infty}[\eta_{\Z^d}]$
is an infinite volume Gibbs measure for $P$-a.e. $\eta$
that depends measurably on $\eta$. Consequently 
the infinite volume joint state is then just the $\P$-integral 
of $\mu_{\infty}$.
We stress that this has not been assumed so far and is 
really a much stronger assumption then local convergence
of the joint states. It is not expected to hold
e.g. for spinglasses in the multi-phase region
(that is supposed although not proved to exist).

This assumption implies that 
the terms in the main formula of Lemma 1 converge individually 
with $\L_N\uparrow\Z^d$. 
So we have that 
$$
\eqalign{
&q^{\hbox{\srm nonloc} }_{\L,x}
[\eta^1_x,\eta^2_x,\eta_{\L\ba x},\s_{\del_{-}\L}]\cr
&=
\int K
\left[d\tilde\eta_{\Z^d\ba\L}\bigl|\s_{\del_{-}\L};\eta^2_{x},\eta_{\L\ba x}
\right]
\mu_{\infty}
[\eta^1_{x},\eta_{\L\ba x},\tilde\eta_{\Z^d\ba \L}]
\left( e^{\D H_{x}(\eta^1_{x},\eta^2_{x},\eta_{\del {x}})}\right)\cr
}
\tag{2.19}
$$
Suppose we want to exhibit a bad configuration and we
have estimates on the continuity 
of $\eta\mapsto\mu_{\infty}[\eta]$ for {\it typical}
directions but {\it not} in all directions.
For an example of a perturbation in an
atypical direction think of the 
random field Ising model that will be discussed below. 
Here the Gibbs-measure with plus boundary conditions
can be pushed in the `wrong phase' by choosing 
the random fields to be minus in a large annulus.
While the RFIM can be treated by Proposition 3
there are examples where we would like to 
get away from {\it uniform} estimates w.r.t. $\tilde\eta$
in favor of estimates that are only true 
for {\it typical} $\tilde\eta$, for the a-priori measure $\P$.

To obtain the following criterion is more subtle
than what we noted in Proposition 2 and 3. 
The trick is to show the existence of suitable 
`bad' $\s$-conditionings using the knowledge about 
typical disorder variables w.r.t. the unbiased 
$\P$-measure.

\bigskip

\proposition{4}{\it
Assume the a.s. existence of the weak limits 
of finite volume Gibbs measures (2.18) and denote
by $K$ the corresponding infinite volume 
joint measure.

The configuration $\xi=(\eta,\s)$ is a bad configuration
for $K$ if: for each cube $V$, centered at the origin, 
there exists an increasing choice of volumes $\L(V)$,
and configurations $\eta^{V},\bar \eta^{V}$ s.t. for 
$\P$-a.e. $\tilde\eta$ we have that 
$$
\eqalign{
&\liminf_{V\uparrow\Z^d}
\mu_{\infty}
[\eta^1_{x},\eta_{V\ba x}\bar\eta^{V}_{\L(V)\ba V},\tilde\eta_{\Z^d\ba \L}]
\left( e^{\D H_{x}(\eta^1_{x},\eta^2_{x},\eta_{\del {x}})}\right)\cr
&>\limsup_{V\uparrow\Z^d}
\mu_{\infty}
[\eta^1_{x},\eta_{V\ba x}\eta^{V}_{\L(V)\ba V},\tilde\eta_{\Z^d\ba \L}]
\left( e^{\D H_{x}(\eta^1_{x},\eta^2_{x},\eta_{\del {x}})}\right)
}
\tag{2.20}
$$
for some site $x$, and some $\eta_x^1,\eta_x^2$.
}

\proof We will show that there exist two conditionings
$\bar \s$ and $\s$, s.t. 
$$
\eqalign{
&\liminf_{V\uparrow\Z^d}
q^{\hbox{\srm nonloc}}_{\L(V),x}[\eta^1_x,\eta^2_x,
\eta_{V\ba x}\bar\eta^{V}_{\L(V)\ba V},\bar\s_{\del_{-}\L(V)}]\cr
&>\limsup_{V\uparrow\Z^d}
q^{\hbox{\srm nonloc}}_{\L(V),x}[\eta^1_x,\eta^2_x,
\eta_{V\ba x}\eta^{V}_{\L(V)\ba V},\s_{\del_{-}\L(V)}]
}
\tag{2.21}
$$
From this and the Corollary of Proposition 1 follows 
the badness.

To show (2.21) we proceed as follows:
The l.h.s. and r.h.s. of (2.20) are tail measurable,
hence a.s. constant. Denote the l.h.s of (2.20) by 
$\bar q^{\infty}[\eta^1_x,\eta^2_x,\eta_{\Z^d\ba x}]$
and the r.h.s. by $q^{\infty}[\eta^1_x,\eta^2_x,\eta_{\Z^d\ba x}]$.
We will show that there exists a conditioning $\s$ s.t. the r.h.s.
of (2.21) is bounded from above by 
$q^{\infty}[\eta^1_x,\eta^2_x,\eta_{\Z^d\ba x}]$. (Similarly, 
there exists a conditioning $\bar \s$
s.t. the l.h.s.
of (2.21) is bounded from below by 
$\bar q^{\infty}[\eta^1_x,\eta^2_x,\eta_{\Z^d\ba x}]$.)

We will construct this conditioning as
a sequence given on the `small' annuli $\del_{-}\L(V)$
(and arbitrary for other lattice sites.)
To make use of the a.s. statement w.r.t the product measure $\P$
we need to produce a formula that recovers this measure.
We write
$$
\eqalign{
&\limsup_{V\uparrow\infty}
\sum_{\tilde\s_{\del_{-}\L(V)}}\int K_{\infty}
\left[\tilde \s_{\del_{-}\L(V)}\bigl |
\eta_{x}^{2},\eta_{V\ba x}\eta^{V}_{\L(V)\ba V}\right]
q^{\hbox{\srm nonloc}}_{\L(V),x}[\eta^1_x,\eta^2_x,
\eta_{V\ba x}\eta^{V}_{\L(V)\ba V},\tilde\s_{\del_{-}\L(V)}]\cr
&=\limsup_{V\uparrow\infty}
\int\P(d\tilde\eta)
\mu_{\infty}
[\eta^1_{x},\eta_{V\ba x}\eta^{V}_{\L(V)\ba V},\tilde\eta_{\Z^d\ba \L}]
\left( e^{\D H_{x}(\eta^1_{x},\eta^2_{x},\eta_{\del {x}})}\right)
\leq q^{\infty}[\eta^1_x,\eta^2_x,\eta_{\Z^d\ba x}]
}
\tag{2.22}
$$
where the first equality follows from (2.19)
and the inequality from Fatou's Lemma  w.r.t product-integration 
of the $\tilde\eta$. 
From this, the existence of such a conditioning
$\s$ is easy to see. (By contradiction:
If the claim were not true, for any sequence
of conditionings $\s_{\del_{-}{\L(V)}}$, 
we would have that there exists a positive $\e$
s.t. $\min_{\tilde\s_{\del_{-}\L(V)}}
q^{\hbox{\srm nonloc}}_{\L(V),x}[\dots,\tilde\s_{\del_{-}\L(V)}] 
\geq q^{\infty}[\dots]+\e$ for infinitely many  $V$'s. 
But this would imply that also the  quantity 
under the limsup on the l.h.s. of (2.22)
[which is just a $\tilde\s_{\del_{-}\L(V)}$-expectation]
would have to be bigger of equal to this bound, for the 
same infinitely many $V$'s.)
\endproof

\bigskip
\bigskip

\chap{III. Examples \hfill}
\medskip

\line{\bf III.1: The random field Ising model \hfill}

Note that the single site perturbation w.r.t the random field
of the Hamiltonian is very simple, i.e.
$$
\eqalign{
&e^{\D H_{x}(\s_{x},\eta_{x}^1,\eta_{x}^2)}= e^{h(\eta^2_x-\eta_x^1)\s_x}
=e^{h(\eta_x^1-\eta_x^2)}+2\sinh h(\eta_x^2-\eta_x^1)\,\, 1_{\s_x=1}
\cr
}
\tag{3.1}
$$
An application of Propositions 2 and 3 gives, 
with the aid of monotonicity arguments Theorem 1, as
stated in the introduction.
It provides a complete characterization 
of good/bad configurations in terms of the behavior
of the finite volume Gibbs-expectations with 
plus resp. minus boundary conditions.  
The interesting part, the mechanism of non-continuity, 
is due to the fact that we can make the random field
Gibbs measure look like the plus (minus) phase around a given site 
by choosing the fields in a sufficiently large 
annulus to be plus (minus). 
That this works independently of what the fields
even further outside do, is crucial for the argument.  


\proofof{Theorem 1} We use the fact that the function 
$(\eta,\s^{\hbox{\srm bc}})\mapsto
\mu^{\s^{\hbox{\srm bc}}}_{\L}
[\eta_{\L}]\left(\tilde \s_x=1\right)$
is monotone (w.r.t. the partial order of its arguments obtained by
site-wise comparison.) 
From this follows that the limits in (1.7) exist, due 
to monotonicity, for any $\eta$.  Denote the 
l.h.s. of (1.7) by 
$m^+_{x}(\eta_{\Z^d})$ and the r.h.s. of (1.7)
by $m^-_{x}(\eta_{\Z^d})$.
We also note that, by the finite-volume perturbation
formula, one obtains that 
$$
\eqalign{
&e^{h(\eta_x^1-\eta_x^2)}
\left(\left[m^{+,-}_{x}(\eta_x^1,\eta_{\Z^d\ba x})\right]^{-1}-1\right)
= e^{h(\eta_x^2-\eta_x^1)}
\left(\left[m^{+,-}_{x}(\eta_x^2,\eta_{\Z^d\ba x})\right]^{-1}-1\right)
\cr
}
\tag{3.2}
$$
This shows in particular that (say) 
$m^{+}_{x}(\eta_x^1,\eta_{\Z^d\ba x})$
and $m^{+}_{x}(\eta_x^2,\eta_{\Z^d\ba x})$ are strictly monotone
functions of each other (when varying
$\eta_{\Z^d\ba x}$).  In particular we see explicitly that, 
whether the l.h.s. and r.h.s. of (1.7) coincide 
does of course not depend on the value of $\eta_x$.  

\medskip
Now, to show that a configuration is good if the two 
limits coincide, we apply Proposition 2 and the remark after it.  
Using (3.1), we see  
that $r_{V,x}(\eta^1_{x},\eta^2_{x},\eta)\rightarrow 0$
with $V\uparrow\Z^d$ if  
$$
\eqalign{
&\sup_{\s^1,\s^2}\left|
\int\mu^{\s^{1}_{\del V}}_{V}
[\eta^1_x\eta_{V\ba x}](\tilde\s_x=1)
- \int\mu^{\s^{2}_{\del V}}_{V}
[\eta^1_x\eta_{V\ba x}](\tilde\s_x=1)\right|
\rightarrow 0
}
\tag{3.3}
$$
with $V\uparrow\Z^d$.
Using monotonicity in the boundary condition
we see that this is equivalent to the equality of the
two limits in (1.7).

\medskip
Now, to show that a configuration is bad, if the 
two limits in (1.7) do not coincide, we use 
Proposition 3. We have that 
$$
\eqalign{
&q^{\hbox{\srm upper} }_{\L,x}[\eta^1_x,\eta^2_x,\eta_{\L\ba x}]=
\limsup_{\L_N\uparrow\Z^d}
\sup_{\tilde \eta_{\L_N\ba \L}}
\mu^{\s^{\hbox{\srm b.c.}}_{\del\L_N}}_{\L_N}
[\eta_{x}^{1},\eta_{\L\ba x},\tilde \eta_{\ov\L_N\ba \L}]
\left(e^{\D H_{x}(\s_{x},\eta_{x}^1,\eta_{x}^2)}\right)
\cr
&=
e^{h(\eta_x^1-\eta_x^2)}+
\limsup_{\L_N\uparrow\Z^d}
\sup_{\tilde \eta_{\L_N\ba \L}} 2\sinh(h(\eta_x^2-\eta_x^1))
\mu^{\s^{\hbox{\srm b.c.}}_{\del\L_N}}_{\L_N}
[\eta_{x}^{1},\eta_{\L\ba x},\tilde \eta_{\ov\L_N\ba \L}]
\left(\tilde \s_x=1\right)
}
\tag{3.4}
$$
Suppose now that $\eta^2_x\geq \eta_x^1$.
Then we get from the monotonicity 
$$
\eqalign{
&q^{\hbox{\srm upper} }_{\L,x}[\eta_x^1,\eta^2_x,\eta_{\L\ba x}]
\leq e^{h(\eta_x^1-\eta_x^2)}+
2\sinh(h(\eta_x^2-\eta_x^1)) 
\mu^{+_{\del\L}}_{\L}[\eta_{x}^{1},\eta_{\L\ba x}]
\left(\tilde \s_x=1\right)\cr
}
\tag{3.5}
$$
Similarly we have that 
$$
\eqalign{
&q^{\hbox{\srm upper} }_{\L,x}[\eta_x^2,\eta^1_x,\eta_{\L\ba x}]
\leq e^{h(\eta_x^2-\eta_x^1)}+
2\sinh(h(\eta_x^1-\eta_x^2)) 
\mu^{-_{\del\L}}_{\L}[\eta_{x}^{1},\eta_{\L\ba x}]
\left(\tilde \s_x=1\right)\cr
}
\tag{3.6}
$$
Now we use the important fact that 
$$
\eqalign{
&\lim_{\L\uparrow \Z^d}
\mu^{-_{\del\L}}_{\L}
[\eta_{V},\eta_{\L\ba V}=+]
\left(\tilde \s_x=1\right)
= \lim_{\L\uparrow \Z^d}
\mu^{+_{\del\L}}_{\L}
[\eta_{V},\eta_{\L\ba V}=+]
\left(\tilde \s_x=1\right)
}
\tag{3.7}
$$
that follows from the unicity of 
the Gibbs measure of a homogeneous ferromagnet
in a positive magnetic field, and, consequently,  
$$
\eqalign{
&\lim_{V\uparrow \Z^d}\lim_{\L\uparrow \Z^d}
\mu^{-_{\del\L}}_{\L}
[\eta_{V},\eta_{\L\ba V}=+]
\left(\tilde \s_x=1\right)
=\lim_{V\uparrow \Z^d} \lim_{\L\uparrow \Z^d}
\mu^{+_{\del\L}}_{\L}
[\eta_{V},\eta_{\L\ba V}=+]
\left(\tilde \s_x=1\right)\cr
&=m^+_{x}(\eta_{\Z^d})
}
\tag{3.8}
$$
where the right equality follows 
from the inequality $\mu^{+_{\del\L}}_{\L}
[\eta_{V},\eta_{\L\ba V}]
\left(\tilde \s_x=1\right)
\leq \mu^{+_{\del\L}}_{\L}
[\eta_{V},\eta_{\L\ba V}=+]
\left(\tilde \s_x=1\right)\leq 
\mu^{+_{\del V}}_{V}
[\eta_{V}]
\left(\tilde \s_x=1\right)$. From this we have that 
$$
\eqalign{
&\lim_{V\uparrow \Z^d}\lim_{\L\uparrow \Z^d}
q^{\hbox{\srm upper} }_{\L,x}[\eta_x^1,\eta^2_x,\eta_{V},\eta_{\L\ba V}=-]
\leq e^{h(\eta_x^1-\eta_x^2)}+
2\sinh(h(\eta_x^2-\eta_x^1)) m^-_{x}(\eta^1_x,\eta_{\Z^d\ba x})\cr
}
\tag{3.9}
$$
and, similarly 
$$
\eqalign{
&\lim_{V\uparrow \Z^d}\lim_{\L\uparrow \Z^d}
q^{\hbox{\srm upper} }_{\L,x}[\eta_x^2,\eta^1_x,\eta_{V},\eta_{\L\ba V}=+]
\leq e^{h(\eta_x^2-\eta_x^1)}+
2\sinh(h(\eta_x^1-\eta_x^2)) m^+_{x}(\eta^2_x,\eta_{\Z^d\ba x})\cr
&=\left( 
e^{h(\eta_x^1-\eta_x^2)}+
2\sinh(h(\eta_x^2-\eta_x^1)) m^+_{x}(\eta^1_x,\eta_{\Z^d\ba x})
\right)^{-1}
}
\tag{3.10}
$$
where the last line follows from relation (3.2).
From this it is evident that (1.7) implies (2.16).
\endproof

\bigskip
\bigskip

\line{\bf III.2: Models with decoupling configurations\hfill}

Suppose we have a model that allows for `non-percolating'
decoupling configurations $\eta$. By this we mean that, 
for given $\eta$, for any site $x$ there exists a volume $\L_x(\eta)$ s.t.,
for any $\L\sp \L_x(\eta)$ we have that 
$$
\eqalign{
&\mu^{\s^{\hbox{\srm b.c.}}_{\del\L}}_{\L}
[\eta^1_{x},\hat\eta_{\ov{\L}\ba x}]\left(
e^{\D H_{x}(\eta^1_{x},\eta^2_{x},\eta_{\del {x}})}\right)
=\mu^{\hbox{\srm open}}_{\L_x(\eta)}[\eta_x^1\eta_{\L_x(\eta)\ba x}]
\left(e^{\D H_{x}(\eta^1_{x},\eta^2_{x},\eta_{\del {x}})}
\right)
}
\tag{3.11}
$$
independently of $\L$ (for any pair $\eta_x^1,\eta_x^2$),
for any configuration $\hat\eta$
that coincides with $\eta$ on $\L_x(\eta)$.

Think e.g. of an Ising model with random couplings  
taking the value $0$ with positive probability.
Then a configuration of coupling constants
s.t. the all resulting spin clusters
(with edges of non-zero coupling constants) are finite
is such a non-percolating decoupling configuration. 

For a decoupling configuration $\eta$ the formula for the conditional
expectations simplifies considerably.
A look at (2.10) tells us that we get 
$$
\eqalign{
&q^{\hbox{\srm nonloc} }_{\L,x}[\eta^1_x,\eta^2_x,\eta_{\L\ba x},\s_{\del_{-}\L}]
=\mu^{\hbox{\srm open}}_{\L_x(\eta)}[\eta_x^1,\eta_{\ov{\L_x(\eta)}\ba x}]
\left(e^{\D H_{x}(\eta^1_{x},\eta^2_{x},\eta_{\del {x}})}
\right)
}
\tag{3.12}
$$
for $\L$ sufficiently large (depending on $\eta$). 
Since any perturbation of $\eta$ far away from $x$ 
leaves this quantity unchanged, we immediately obtain:

\proposition{5}{\it 
A configuration $\x=(\eta,\s)$ is a good configuration,
if $\eta$ is a decoupling configuration.  
Consequently:  
If $\P\left[
\eta\in \HH: \eta \hbox{ is a decoupling configuration}
\right]=1$, then any joint measure that is a limit of the 
form (1.3) is almost surely Gibbs.
}

This has not to be confused with the fact that a non-decoupling
$\eta$ can be shown to be bad with
the use of (a sequence of) decoupling configurations $\eta^+$, $\eta^{-}$,
as the following examples show. 

\bigskip

\line{\bf The GriSing Random Field revisited (see [EMSS]):\hfill}

The spins are $\s_x\in \{-1,1\}$, the local disorder variable
$\eta_x$ takes values in $\{0,1\}$ with $\nu[\eta_x=1]=p\in (0,1)$ 
and the Hamiltonian 
is given by $H^{\eta}(\s)=-J\sum_{<x,y>}\eta_x\s_x\eta_y\s_y$.
This model was shown to be non-Gibbs
for $p$ below the percolation threshold for site percolation.  
Let us see, how this comes out of our framework 
and explain at the same time that {\it any}\footnote{$^1$}{Think e.g.
of the Dobrushin states that are supposed to exist
for $p$ close to $1$ in $d\geq 4$!} weak 
limit $\lim_{\L_N}
\P(d\eta)\mu_{\L_N}^{\s^{\hbox{\srm b.c.}}}[\eta](d\s)$
will also be non-Gibbs, for any $p\in (0,1)$ 
(for sufficiently large $J$).

There is the trivial mapping that sends the pair $(\eta_x,\s_x)$
to the product $\eta_x\s_x$; 
looking at new variables that are 
products (as it was done in [EMSS]) is equivalent
to looking at pairs 
since $\eta_x=0$ iff $\eta_x\s_x=0$. 

Recalling [EMSS] we
look at the configuration $\eta^{\srm disc}$ 
that is $0$ on the `base-plane' $B=\{x\in \Z^d, x_d=0\}$
and $1$ otherwise. Then $(\eta^{\srm disc},\s)$ is a bad 
configuration for any $\s$, for any joint infinite volume
measure that is a limit of the form (1.3).  
To see this, one only needs to look at conditional
probabilities for special decoupling configurations. 
Indeed, for a finite box $V\sb \Z^d$, centered 
at the origin, denote by $\eta^{\srm disc,V}$
the configuration that coincides with $\eta^{\srm disc}$
inside $V$ and vanishes outside $V$. 
Denote by $V^+$ ($V^-$)
the occupied sites in $V$ in the upper (lower) half-space.  
For $z\in V\cap B$ denote by $\eta^{\srm disc,V,z}$ 
the configuration
that has $z$ as an additional occupied site. 
Denote the nearest neighbor of the origin in $V^+$
by $x_0$ and the nearest neighbor of the origin in $V^-$
by $y_0$. 
Put $\eta_0^2=1$, $\eta_0^1=0$.
Then 
$e^{\D H_{0}(\s_{\ov 0},\eta_{0}^1,\eta_{0}^2,\eta^{\srm disc}_{\del 0})}= 
e^{J\s_0(\s_{x_0} +\s_{y_0})}$ and one obtains
$$
\eqalign{
&q^{\hbox{\srm nonloc} }_{\L,0}[\eta^1_0,\eta^2_0,\eta^{\srm disc,V}
_{\L\ba 0},
\s_{\del_{-}\L}]
= 2 \mu_{V^+\cup V^-}^0 \left( \cosh J(\tilde \s_{x_0} +\tilde 
\s_{y_0})\right)= a \mu_{V^+\cup V^-}^0
\left(\tilde \s_{x_0}\tilde \s_{y_0}\right) 
+b \cr
&q^{\hbox{\srm nonloc} }_{\L,0}[\eta^1_0,\eta^2_0,\eta^{\srm disc,V,z}
_{\L\ba 0},
\s_{\del_{-}\L}]
=a \mu_{V^+\cup V^- \cup z}^0
\left(\tilde \s_{x_0}\tilde \s_{y_0}\right)+b 
}
\tag{3.13}
$$
for some positive constants $a,b$, for $\L$ sufficiently large.
Here $\mu^0_{W}$ is the ferromagnetic Ising Gibbs measure
in the finite volume $W$ with zero boundary conditions. 
The correlations on the r.h.s. were seen in [EMSS] to be 
different for large $J$, 
for arbitrarily large $V$, uniformly in the location
of $z$.  (Adding a site $z$ destroys the independence
and introduces a positive correlation between 
$\s_{x_0}$ and $\s_{y_0}$ once there is ferromagnetic order.)
By the Corollary of Proposition 1 this shows
that $(\eta^{\srm disc},\s)$ is a bad 
configuration for any $\s$.  

Our point here was that while $\eta^{\srm disc}$ 
is not a decoupling configuration, 
the perturbed configurations
$\eta^{\srm disc,V},\eta^{\srm disc,V,z}$ 
are decoupling, leading to simple formulas for $q^{\srm nonloc}$,
that are independent of the specific joint measure and independent
of the value of $p$.

\bigskip

\line{\bf Models with Random Bonds that can be zero:\hfill}

We note that the same [EMMS]-mechanism is responsible
for the occurrence of bad configurations in models
with random bonds. Although not difficult to see once 
the previous example is understood, 
this might be interesting, because it is also 
true for e.g. for EA 
spinglasses of the type (iib) from the Introduction. We have 

\proposition{6}{\it Suppose that we are given a model of the form 
(1.5) in dimensions $d\geq 2$ 
where $\nu(J_{x,e}=0)>0$ and $\nu(J_{x,e}=J^1)>0$
with $J^1$ sufficiently large. 

Decompose the lattice $\Z^d$ into two half-spaces
$\Z^d_+\cup \Z^d_{-}$ that are separated by a
hyper-plane of bonds that we call $H$. 
Denote by  $J^{\srm disc}$ the configuration of bonds
that is equal to zero for bands in $H$ and equal  
to $J^1$ otherwise. 

Then $\xi=(J^{\srm disc},\s)$ is a bad configuration
for any joint measure obtained as limit point of  $\P(d J)
\mu^{\s^{\hbox{\srm b.c.}}_{\del\L}}_{\L}[J](d\s)$.
}

\proof Assume that the hyper-plane is of the form
$H=\{<x,y>: x_d=0, y_d=1\}$. Then we have $<0,e_d>\in H$. 
In a similar fashion 
as above, for finite $\tilde V\sb (\Z^d)^*$ (a box on the dual
lattice, centered around the origin), denote by $J^{\srm disc,\tilde 
V}$ the configuration that coincides with $J^{\srm disc}$
inside $\tilde V$ and vanishes outside $\tilde V$. 
Denote by $\tilde V^+$ ($\tilde V^-$)
the occupied bonds in $\tilde V$ in the upper (lower) half-space.  
For a bond $b\in \tilde V\cap H$ denote by $J^{\srm disc,\tilde V,b}$ 
the configuration
that has $b$ as an additional non-empty coupling taking 
the value $J^1$.

To find a discontinuity, it 
suffices to look at pairs 
$\eta_x^1$ and $\eta_x^2$ that differ
only by one coupling constant, 
$\eta^1_0=(J^1_{0,e_1},\dots,J^1_{x,e_{d-1}},0)$  
and 
$\eta^2_0=(J^1_{0,e_1},\dots,J^1_{x,e_{d-1}},J^1_{x,e_{d}} )$.
Then the  variation at the origin becomes
$e^{\D H_{0}(\s_{\ov 0},\eta_{0}^1,\eta_{0}^2, \eta^{\srm disc}_{\del 0})}= 
e^{J^1\s_{0}\s_{e_d}}
=e^{-J^1}+2\sinh J^1\,\, 1_{\s_0=\s_{e_d}}$
where we have written $\eta^{\srm disc}$ for  
the obvious configuration corresponding to $J^{\srm disc}$ (and will
also do so for $\eta^{\srm disc,\tilde V}$, $\eta^{\srm disc,\tilde V,b}$).
So one obtains
$$
\eqalign{
&q^{\hbox{\srm nonloc} }_{\L,0}[\eta^1_0,\eta^2_0,\eta^{\srm disc,\tilde V}
_{\L\ba 0},
s_{\del_{-}\L}]
= e^{-J^1}+2\sinh J^1\,\,\hat\mu_{\tilde V^+\cup \tilde V^-}^0 \left(
 \tilde\s_0=\tilde \s_{e_d}\right)\cr
&q^{\hbox{\srm nonloc} }_{\L,0}[\eta^1_0,\eta^2_0,\eta^{\srm disc,\tilde V,b}
_{\L\ba 0},
\s_{\del_{-}\L}]
= e^{-J^1}+2\sinh J^1\,\,\hat\mu_{\tilde V^+\cup \tilde V^-\cup b}^0 \left(
 \tilde\s_0=\tilde \s_{e_d}\right)\cr
}
\tag{3.14}
$$
for $\L$ sufficiently large.
Here $\hat\mu^0_{\tilde W}$ is the ferromagnetic Ising Gibbs measure
with zero boundary conditions
on the vertex set of the graph whose bonds are $\tilde W$
with the coupling constant $J^1$. 
Now, in the very same way as in [EMSS], 
the probabilities  on the r.h.s.'s are seen to be 
different, 
for arbitrarily large $\tilde V$, uniformly in the location
of $b$. By the Corollary of Proposition 1 this shows
the claim. 
\endproof

\bigskip
\bigskip

\line{\bf III.3: Ising models with disordered nearest neighbor
couplings \hfill}

Denote by $(\Z^d)^*$ the lattice of bonds of $\Z^d$.
We denote subsets of $(\Z^d)^*$ by symbols with tildes
(like $\tilde V$) . 
An application of Proposition (2) resp. Proposition (4)
yields the following.

\theo{2}{\it Consider an Ising model with 
random nearest neighbor couplings of the form (1.5), in any dimension $d$.

\item{(i)} A configuration $\x=( J,\s)$
is a {\bf good} configuration for any joint measure
obtained as a limit point of the finite volume 
joint measures $\P(d J)
\mu^{\s^{\hbox{\srm b.c.}}_{\del\L}}_{\L}[J](d\s)$ if 
$$
\eqalign{
&\sup_{{J^+,J^-}\atop{\L}}
\Bigl|\mu^{\s^{\hbox{\srm b.c.}}_{\del \L}}_{\L}
[J_{\tilde V}J^+_{(\Z^d)^*\ba \tilde V}]
(\tilde\s_{x}=\tilde\s_{y})
-\mu^{\s^{\hbox{\srm b.c.}}_{\del \L}}_{\L}
[J_{\tilde V}J^-_{(\Z^d)^*\ba \tilde V}]
(\tilde\s_{x}=\tilde\s_{y})
\Bigr|\rightarrow 0
}
\tag{3.15}
$$
with $\tilde V\uparrow (\Z^d)^*$.

\item{(ii)} Suppose moreover 
that we have the existence of a weak limit 
$\lim_{\L\uparrow\Z^d}
\mu^{\s^{\hbox{\srm b.c.}}_{\del\L}}_{\L}[J]
= \mu_{\infty}[J]$
for a nonrandom boundary condition $\s^{\hbox{\srm b.c.}}$,
for $\P$-a.e. $J$. Denote by $K(d\s,d J)= 
\P(d J)\mu_{\infty}[J](d\s)$ the corresponding joint 
measure. 

\item{} A configuration $\x=(J,\s)$
is a {\bf bad} configuration for $K$, if there exists
an increasing choice of volumes $\tilde\L(\tilde V)$ and 
configurations $J^{\tilde V},\bar J^{\tilde V}$, 
s.t., for  $\P$-a.e. $\tilde J$ 
we have that
$$
\eqalign{
&\liminf_{\tilde V\uparrow(\Z^d)^*}
\mu_{\infty}[J_{\tilde V} 
\bar J^{\tilde V}_{\tilde \L(\tilde V)\ba \tilde V},
\tilde J_{(\Z^d)^*\ba \tilde \L(V)}](\tilde \s_x
=\tilde\s_y) \cr
&>\limsup_{\tilde V\uparrow(\Z^d)^*}
\mu_{\infty}[J_{\tilde V} 
J^{\tilde V}_{\tilde \L(\tilde V)\ba \tilde V},
\tilde J_{(\Z^d)^*\ba \tilde \L(V)}](\tilde \s_x
=\tilde\s_y)
}
\tag{3.16}
$$
for some nearest neighbor pair $<x,y>$.

}

\proof
To check the condition of Proposition 2, it 
suffices to look at pairs 
$\eta_x^1$ and $\eta_x^2$ that differ
only by one coupling constant, say 
$\eta^1_x=(J_{x,e_1},\dots,J_{x,e_{j-1}},J^1,J_{x,e_{j+1}},
\dots,J_{x,e_d} )$
and $\eta^2_x=(J_{x,e_1},\dots,J_{x,e_{j-1}},J^2,J_{x,e_{j+1}},
\dots,J_{x,e_d})$.
Put $y=x+e_{j}$.
The variations at the site $x$ then become
$$
\eqalign{
&e^{\D H_{x}(\s_{\ov x},\eta_{x}^1,\eta_{x}^2)}= 
e^{(J^2-J^1)\s_x\s_y}
=e^{(J^1-J^2)}+2\sinh (J^2-J^1)\,\, 1_{\s_x=\s_y}\cr
}
\tag{3.17}
$$
which is analogous to formula (3.1) for the Random field model. 

Writing out the condition (2.12) from Proposition 2
then essentially amounts 
to the criterion (3.15) given in the theorem, 
except that possibly different values $J$ at the bond $<x,y>$
can appear. However, there is a simple formula analogous
to formula (3.2) for the random field model
relating the probabilities of the event $\s_x=\s_y$
for different values of $J_{<x,y>}$ that is obtained
by the finite volume perturbation formula. 
From this an argument like the one given for the 
random field model given after (3.2) shows that the validity
of condition (3.15) is independent of the value of $J_{<x,y>}$.
This proves statement (i).

To show that $(J,\s)$ is a bad configuration (for any $\s$)
by means of Proposition 4 we have to look at 
$$
\eqalign{
&\limsup_{\tilde V\uparrow(\Z^d)^*}
\mu_{\infty}[J^1_{<x,y>},J_{\tilde V\ba <x,y>} 
J^{\tilde V}_{\tilde \L(\tilde V)\ba \tilde V},
\tilde J_{(\Z^d)^*\ba \tilde \L(V)}](e^{(J^2-J^1)\tilde \s_x
\tilde\s_y})\text{and}\cr 
&\liminf_{\tilde V\uparrow(\Z^d)^*}
\mu_{\infty}[J^1_{<x,y>},J_{\tilde V\ba <x,y>} 
\bar J^{\tilde V}_{\tilde \L(\tilde V)\ba \tilde V},
\tilde J_{(\Z^d)^*\ba \tilde \L(V)}](e^{(J^2-J^1)\tilde \s_x
\tilde\s_y}) \cr
}
\tag{3.18}
$$
and find two sequences of conditionings $J^{\tilde V}$
and $\bar J^{\tilde V}$ such that the lower expression
is strictly bigger than the upper one. 
Assuming that $J^2>J^1$, this is true, if and only if 
$$
\eqalign{
&\liminf_{\tilde V\uparrow(\Z^d)^*}
\mu_{\infty}[J^1_{<x,y>},J_{\tilde V\ba <x,y>} 
\bar J^{\tilde V}_{\tilde \L(\tilde V)\ba \tilde V},
\tilde J_{(\Z^d)^*\ba \tilde \L(V)}](\tilde \s_x
=\tilde\s_y) \cr
&>\limsup_{\tilde V\uparrow(\Z^d)^*}
\mu_{\infty}[J^1_{<x,y>},J_{\tilde V\ba <x,y>} 
J^{\tilde V}_{\tilde \L(\tilde V)\ba \tilde V},
\tilde J_{(\Z^d)^*\ba \tilde \L(V)}](\tilde \s_x
=\tilde\s_y)
}
\tag{3.19}
$$
Using the argument presented for the RFIM we see that this is true 
if and only if the same strict inequality holds
for any other value of $J_{<x,y>}$ replacing $J^1_{<x,y>}$.
This proves statement (ii).
\endproof

\bigskip

Finally we would like to discuss the relevance of Theorem 2
on a heuristic level in application to a random bond ferromagnet. 
\medskip

\line{\bf Heuristics considerations: Gibbsianness destroyed by interfaces\hfill}

Assume dimensions $d\geq 2$.
Suppose that the random bonds $J_{x,e}$ take two values
$0<J^1<J^2<\infty$ with positive probability, independently 
of the bond $(x,e)$. 
We assume that $J^1$ is smaller than the critical inverse temperature
of the corresponding homogeneous Ising ferromagnet.
$J^2$ should be large enough 
and $\nu[J_{x,e}=J^1]$ should be small enough s.t.
there is ferromagnetic order in the disordered model with $\P$-
probability one. 

Let us at first look at $\s^{\hbox{\srm b.c.}}_x\equiv 1$ boundary conditions.
Then we expect a.s. joint Gibbsianness. 
Indeed, Criterion (i) of Theorem 2 should be satisfied
for $\P$-a.e. configuration of couplings $J$, for the following reason:

Let us assume that the realization $J$ is from the full measure
set of couplings for which the finite volume 
Gibbs-measures converge to a ferromagnetic infinite volume 
Gibbs measure. 
Let us check the expected behavior with two `extreme' choices of perturbations:

Consider first a typical perturbation $J^+$ that does have enough
stronger couplings to support the ferromagnetic order. Then 
the state $\mu^{+}_{\L}[J_{\tilde V}J^+_{(\Z^d)^*\ba \tilde V}]$
should look like $\mu^+_{\infty}[J]$ locally, 
for sufficiently large inner volume $\tilde V$ and any (bigger) $\L$.

Choosing next $J^+\equiv J^1$ (the weaker couplings) 
will however destroy the ferromagnetic order in the annulus.
Hence the boundary conditions should be forgotten for 
sufficiently large annulus
and the volume $V$ will approximately feel open boundary conditions. 
(This argument is of course strictly true for the case $J^1=0$). 
The corresponding state should then approximately look like 
$\frac{1}{2}\left(\mu^{+}_{\infty} [J]
+\mu^{-}_{\infty} [J]\right)$ for large $\tilde V$.
This would of course lead to different expectations 
on general observables compared to those of $\mu^{+}_{\infty} [J]$.
The point is however that the expectations of the different states 
on the event $\{\s_x=\s_y\}$ are the same, due to spin-flip symmetry. 

We expect that in general, choosing whatever annulus should 
result in one of the two possibilities, or a linear combination
of them. 

This provides an example that shows 
that although a phase transition occurs by varying
the disorder variables in a large annulus, 
it leads to the same expectations on 
the single site perturbation of the Hamiltonian.
Thus the resulting state can be Gibbs.

\bigskip

Let us now look at Dobrushin boundary conditions,
i.e. we start from finite volume Gibbs measures
in boxes centered around the origin 
with plus boundary conditions on the top 
half, and minus boundary condition on the lower half. 
We assume additionally that we are in dimensions $d\geq 4$,
that $J^2$ is large enough and $\nu[J_{x,e}=J^1]$ small enough s.t.
there are interface states (random `Dobrushin'-states [Do1]) in the disordered model 
with $\P$-probability one. The existence of such states
that are perturbations of the spin configuration that is all plus 
in the upper half-space and all minus in the lower half-space
was proved in [BoK1] in the SOS-approximation of the model. 
(For complementary 
information about disordered interface models, see [BoK2], [K7].) 

Now we expect almost sure non-Gibbsianness
for the resulting infinite volume joint measure, 
different from the model with $+$-boundary conditions.
Indeed, Criterion (ii) of Theorem 2 should be satisfied
for $\P$-a.e. configuration of couplings $J$, for the following reason:

We fix a nearest neighbor pair $<x,y>$ located at, and perpendicular to, 
the base plane (whose intersection
with the boundary of $\L$ is the boundary 
between plus and minus boundary spins).
Again we look first at a typical perturbation $J^+$.
We expect that the infinite volume Dobrushin states 
$\mu^{\pm}_{\infty}[J]$ have the locality property 
that for  $\P$-a.e. perturbation $\tilde J$ 
we have that
$$
\eqalign{
&\lim_{\tilde V\uparrow(\Z^d)^*}
\mu_{\infty}^{\pm}[J_{\tilde \L}
\tilde J_{(\Z^d)^*\ba \tilde \L}](\tilde \s_x
=\tilde\s_y) = 
\mu_{\infty}^{\pm}[J_{(\Z^d)^*}](\tilde \s_x
=\tilde\s_y)
}
\tag{3.20}
$$
for any nearest neighbor pair $<x,y>$.
A corresponding statement 
could in principle be extracted from 
the renormalization group analysis of [BoK1] for
the corresponding SOS-model.

Choosing next the exceptional 
configuration $J^+\equiv J^1$
in an annulus $\tilde \L(\tilde V)\ba \tilde V$
that is sufficiently large will destroy the ferromagnetic order 
in the annulus and decouple the volume $\tilde V$ from the outside. 
This should result in  
$$
\eqalign{
&\lim_{\tilde V\uparrow(\Z^d)^*}
\mu^{\pm}_{\infty}[J_{\tilde V} 
J^{1}_{\tilde \L(\tilde V)\ba \tilde V},
\tilde J_{(\Z^d)^*\ba \tilde \L(V)}](\tilde \s_x
=\tilde\s_y) \cr
&=\frac{1}{2}
\left(
\mu_{\infty}^{+}[J_{(\Z^d)^*}](\tilde \s_x
=\tilde\s_y)
+\mu_{\infty}^{-}[J_{(\Z^d)^*}](\tilde \s_x
=\tilde\s_y)
\right)
}
\tag{3.21}
$$
Note that both terms of the r.h.s. are the same, 
due to spin-flip symmetry. 
This will differ from the expectation in the interface-state (3.20), 
so that we believe that criterion (3.16) should be satisfied.

Let us point out that, in order to reach this 
conclusion even on the heuristic level we have presented it, we really 
needed Theorem 2 (ii) that follows from 
Proposition 4, a result that involves typical configurations
(as opposed to the Criterion of Proposition 3, a result that involves
uniform estimates).
Note that  there is the following fundamental difference
between the random field and the random bond Ising model: 
In the random field model, one 
is able to select a phase by choosing 
the disorder variables (magnetic fields) in a large annulus, no
matter what the disorder variables even further outside will look like. 
In contrast to that, one is not able to `restore' a Dobrushin state
in a random bond model 
by a suitable choice of $J$'s in a large annulus, if the 
$\pm$ boundary conditions have been forgotten, because the couplings
further outside were too weak. 
\bigskip

\bigskip
\bigskip


\ftn
\font\bf=cmbx8

\baselineskip=10pt
\parskip=4pt
\rightskip=0.5truecm
\bigskip\bigskip
\chap{References}
\medskip


\item{[AW]} M.Aizenman, J.Wehr, Rounding Effects of Quenched Randomness
on First-Order Phase Transitions, Comm. Math.Phys {\bf  130},
489-528 (1990)








\item{[BK]} J.Bricmont, A.Kupiainen, 
Phase transition in the 3d random field Ising model,
Comm.
Math.Phys. {\bf 142}, 539-572 (1988)


\item{[BKL]} J.Bricmont, A.Kupiainen, R. Lefevere,
       Renormalization Group Pathologies and the 
Definition of Gibbs States,
Comm. Math.Phys. {\bf 194} 2, 359-388 (1998)
    

\item{[BoK1]} A.Bovier, C.K\"ulske, A rigorous
renormalization group method for interfaces in random
media, Rev.Math.Phys. {\bf 6}, no.3, 413-496 (1994)

\item{[BoK2]} A.Bovier, C.K\"ulske, 
There are no nice interfaces in $2+1$ dimensional 
SOS-models in random media, J.Stat.Phys. {\bf 83}, 751-759
(1996)




\item{[Do1]} R.L.Dobrushin, 
Gibbs states describing a coexistence of phases 
for the three-dimensional Ising model,
Th.Prob. and its Appl. {\bf 17}, 582-600 (1972)

\item{[Do2]} R.L.Dobrushin, 
Lecture given at the workshop `Probability 
and Physics', Renkum, August 1995

\item{[DS]} R.L.Dobrushin, S.B.Shlosman, 
"Non-Gibbsian" states and their Gibbs description, Comm.Math.Phys. 
{\bf 200}, no.1, 125--179 (1999)





\item{[E]} A.C.D.van Enter, The Renormalization-Group peculiarities 
of Griffiths and Pearce: What have we learned? (53K, latex) Oct 30,
available as preprint 98-692  at http://www.ma.utexas.edu/mp\_arc

\item{[ES]} A.C.D.van Enter, S.B.Shlosman,
(Almost) Gibbsian description of the sign fields of
SOS fields. J.Stat.Phys. {\bf 92}, no. 3-4, 353--368 (1998)

\item{[EFS]} A.C.D.van Enter, R. Fern\'andez, A.Sokal,
Regularity properties and pathologies of position-space
renormalization-group transformations: Scope
and limitations of Gibbsian theory. J.Stat.Phys.
{\bf 72}, 879-1167 (1993)

\item{[EMMS]} A.C.D.van Enter, C.Maes, 
R.H.Schonman, S.Shlosman, 
The Griffiths Singularity Random Field,
available as preprint 98-764  at http://www.ma.utexas.edu/mp\_arc
(1998)

\item{[F]} R. Fernandez, Measures for lattice systems,
Physica A {\bf 263} (Invited papers from Statphys 20, Paris (1998)),
117-130 (1999), 
also available as preprint 98-567 at http://www.ma.utexas.edu/mp\_arc


\item{[Geo]} H.O. Georgii, Gibbs measures and phase transitions, Studies
in mathematics, vol. 9 (de Gruyter, Berlin, New York, 1988)


\item{[K1]} C.K\"ulske, Ph.D. Thesis, Ruhr-Universit\"at Bochum (1993)

\item{[K2]} C.K\"ulske, Metastates in Disordered Mean-Field Models:
Random Field and Hopfield Models,
J.Stat.Phys. {\bf 88} 5/6, 1257-1293 (1997)

\item{[K3]} C.K\"ulske, Limiting behavior of random Gibbs measures: metastates 
in some disordered mean field
models, in: 
Mathematical aspects of spin glasses and neural networks,  
Progr. Probab. {\bf 41}, 151-160,
eds. A.Bovier, P.Picco, Birkh\"auser Boston,
Boston (1998)

\item{[K4]} C.K\"ulske, Metastates in Disordered Mean-Field Models II:
The Superstates,
J.Stat.Phys. {\bf 91} 1/2, 155-176 (1998)

\item{[K5]} C.K\"ulske, A random energy model for size dependence:
recurrence vs. transience, Prob.Theor.
Rel.Fields {\bf 111}, 57-100 (1998)

\item{[K6]} 
C.K\"ulske, The continuous spin 
random field model: Ferromagnetic ordering in $d\geq 3$,
to be published in Rev.Math.Phys, available at 
http://www.ma.utexas.edu/mp\_arc/, preprint 98-175 (1998) 

\item{[K7]} C.K\"ulske, Stability for a continuous SOS-interface
model in a randomly perturbed periodic potential, 
available at 
http://www.ma.utexas.edu/mp\_arc/, preprint 98-768 (1998)


\item{[MRM]} C.Maes, F.Redig, A.Van Moffaert, 
     Almost Gibbsian versus Weakly Gibbsian measures,
Stoch.Proc.Appl. {\bf 79}  no. 1, 1--15 (1999), 
also available at 
http://www.ma.utexas.edu/mp\_arc/, preprint 98-193



\item{[N]} C.M.Newman, 
Topics in disordered systems, Lectures in Mathematics ETH Zürich.
Birkh\"auser Verlag, Basel, (1997)

\item{[NS1]} C.M.Newman, D.L.Stein, Spatial Inhomogeneity and
thermodynamic chaos, Phys.Rev.Lett. {\bf 76}, No 25, 4821 (1996) 

\item{[NS2]} C.M.Newman, D.L.Stein, Metastate approach to thermodynamic chaos.,  
Phys. Rev. E {\bf 3} 55, no. 5, part A, 5194-5211 (1997)

\item{[NS3]} C.M.Newman, D.L.Stein, 
Simplicity of state and overlap structure in finite-volume realistic
spin glasses, Phys.Rev.E {\bf 3} 57, no. 2, part A, 1356-1366 (1998)


\item{[NS4]} C.M.Newman, D.L.Stein, Thermodynamic chaos and the structure of short-range
spin glasses, in: Mathematical aspects of spin glasses and neural networks, 243-287, 
Progr. Probab., 41, Bovier, Picco (Eds.), Birkhäuser
Boston, Boston, MA  (1998)


\item{[Se]} T. Sepp\"al\"ainen, Entropy, limit theorems, 
and variational principles for
disordered lattice systems, Commun.Math.Phys {\bf 171},233-277 (1995)


\item{[S]} R.H.Schonmann, Projections of Gibbs measures may
be non-Gibbsian, Comm.Math.Phys. {\bf 124}
1-7 (1989)





\end